%% file: main.tex
\documentclass[aps,twocolumn,superscriptaddress,citeautoscript,floatfix,showpacs,longbibliography]{revtex4-1}

\pdfoutput=1
\usepackage[english]{babel}
\usepackage[utf8]{inputenc}
\usepackage{amsmath}
\usepackage{graphicx}
\usepackage[dvipsnames]{xcolor}
\usepackage{amssymb}
\usepackage{subfigure}
\usepackage{anysize}
\usepackage{physics}
\usepackage{makeidx}
\usepackage{todonotes}
\usepackage{float}
\usepackage{xcolor,colortbl}
\usepackage{listings}
\usepackage{enumitem}
\usepackage[group-separator={,}]{siunitx}

\usepackage[paperwidth=210mm,paperheight=297mm,centering,hmargin=2cm,vmargin=2.5cm]{geometry}

\graphicspath{{images/}}

\definecolor{darkblue}{rgb}{0.1,0.2,0.6}
\definecolor{darkred}{rgb}{0.8,0.1,0.2}
\definecolor{Gray}{gray}{0.9}
\usepackage[colorlinks,citecolor=darkblue,linkcolor=darkred,urlcolor=darkblue]{hyperref}	
\usepackage[all]{hypcap} % fix figure link problem. Links jump to top of figure now, not caption
\providecommand{\newoperator}[2]{\newcommand*{#1}{\mathop{\mathrm{#2}}\nolimits}}
\newoperator{\sgn}{sgn}
\newoperator{\arctanh}{arctanh}
\newoperator{\argmax}{argmax}
\newoperator{\diag}{diag}

\begin{document}

\title{Characterizing the many-body localization transition through correlations}

\def\urbana{
	Institute for Condensed Matter Theory and IQUIST and Department of Physics, 
	University of Illinois at Urbana-Champaign, Urbana, IL 61801, USA
}

\author{Benjamin Villalonga}
	\email{vlllngc2@illinois.edu}
	\affiliation{\urbana}
    
\author{Bryan K. Clark}
	\email{bkclark@illinois.edu}
	\affiliation{\urbana}

\date{\today}

\begin{abstract}
Closed, interacting, quantum systems have the potential to transition to a many-body localized (MBL) phase under the presence of sufficiently strong disorder, hence breaking ergodicity and failing to thermalize.
In this work we study the
distribution of
correlations throughout the ergodic-MBL phase diagram. 
We find the typical correlations in the MBL phase decay as a stretched exponential with range $r$ eventually crossing over to an exponential decay deep in the MBL phase.
At the transition, the stretched exponential goes as $e^{-A\sqrt{r}}$, a decay that is reminiscent of the random singlet phase.
While the standard deviation of the $\log(QMI)$ has a range dependence, the $\log(QMI)$ converges to a range-invariant distribution on all other moments (\emph{i.e.}, the skewness and higher) at the transition.
The universal nature of these distributions provides distinct phenomenology of the transition different from both the ergodic and MBL phenomenologies.
In addition to the typical correlations, we study the extreme correlations in the system, finding that the probability of strong long-range correlations is maximal at the transition, suggesting the proliferation of resonances there.
Finally, we analyze the probability that a single bit of information is shared across two halves of a system, finding that this probability is non-zero deep in the MBL phase but vanishes at moderate disorder well above the transition.

\end{abstract}


\maketitle

\input{introduction.tex}

\input{typical_correlations.tex}

\input{extreme_correlations.tex}

\input{extreme_entanglement.tex}

\input{conclusion.tex}

\input{acknowledgments.tex}

\bibliography{main}

\clearpage

\appendix
\renewcommand\thefigure{A\arabic{figure}}    
\setcounter{figure}{0}    

\input{appendix.tex}

\end{document}

%% file: introduction.tex
\section{Introduction}
\label{sec:intro}

While most phases of matter
are related to the properties of the ground state or the thermal density matrix of a system, eigenstate phases of matter are characterized by properties of a system's interior eigenstates.
The two most well-known eigenstate phases are the many-body localized (MBL) phase, present in sufficiently disordered interacting systems, and the standard ergodic phase into which it transitions~\cite{fleishman_interactions_1980,basko_metalinsulator_2006,gornyi_interacting_2005,rigol_thermalization_2008,nandkishore_many-body_2015,luitz_many-body_2015}.

\begin{figure}[t]
\centering
\includegraphics[width=1.00\columnwidth]{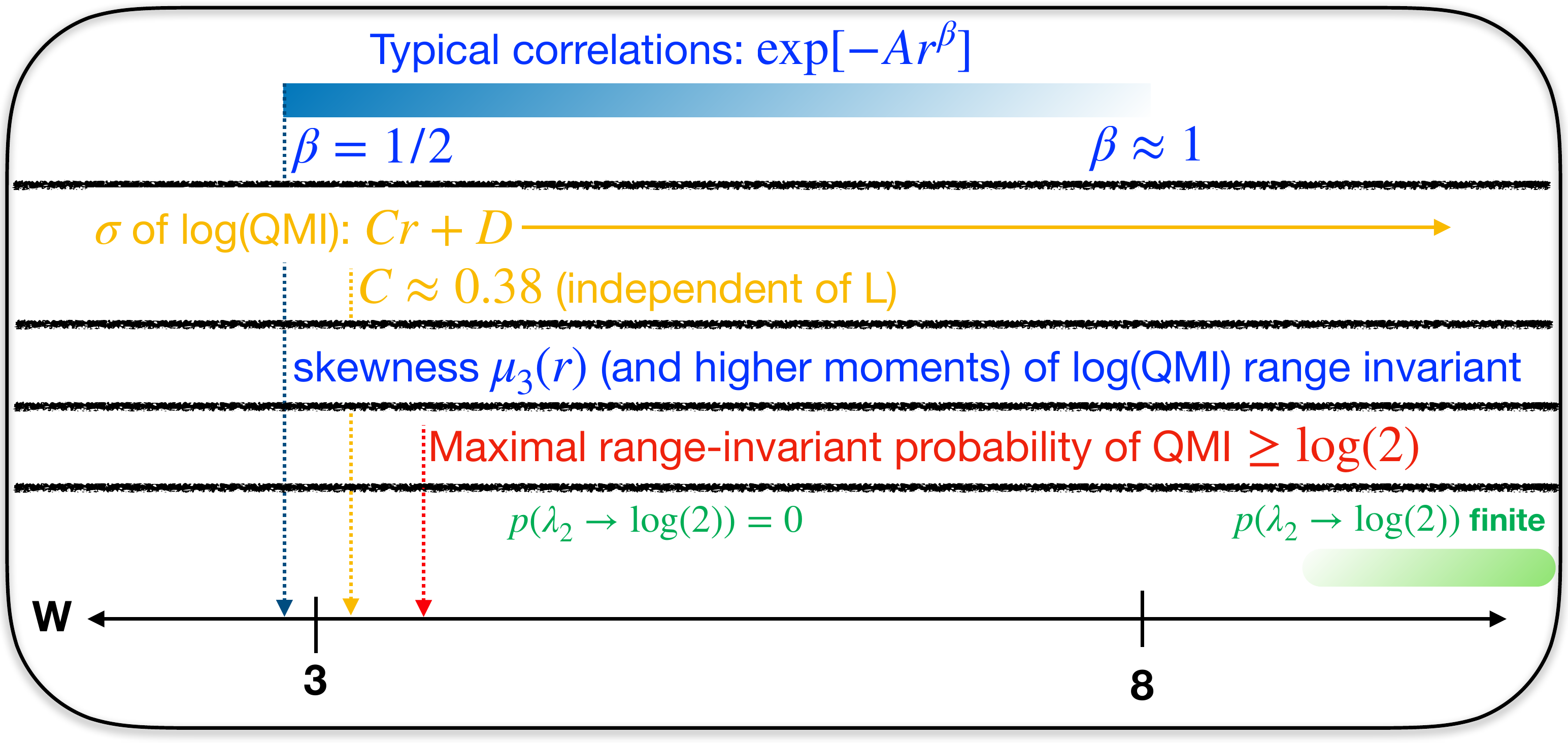}
\caption{\label{fig:summary} Summary of results of this work on the distribution of $\log(QMI)$ and the second Schmidt eigenvalue ($\lambda_2$) of the half-cut reduced density matrix.
Blue bar shows the region of stretched exponential decay as a function of range $r$ ($\exp[-Ar^\beta]$) of the typical $QMI$ (equivalently polynomial decay of the mean of $\log(QMI)$) starting at $\beta=1/2$ at $W_{1 / 2}=2.9$ and going to $\beta \approx 1$ at $W_1 \sim 8$.
The standard deviation of $\log(QMI)$ (yellow) is linear at all $W$ in the MBL phase with an $L$-independent crossing of $C$ as marked on the figure.
The range-invariance of the skewness (and higher moments) of $\log(QMI)$ also happens at $W = W_{1 / 2} = 2.9$.
The probability of finding range-invariant strong values of the $QMI \geq \log 2$ is largest at the $W$ shown, suggesting the proliferation of multi-site resonances.
The green bar indicates the region where the probability of $\lambda_2 = \log 2$ is finite.
All values are at finite $L = 18$ and we expect that $2.9 < W < 4$ all lie within the critical transition region for this $L$, and likely trend towards the same value of $W$ in the thermodynamic limit. 
} 
\end{figure}

\begin{figure}[t]
\centering
\includegraphics[width=1.00\columnwidth]{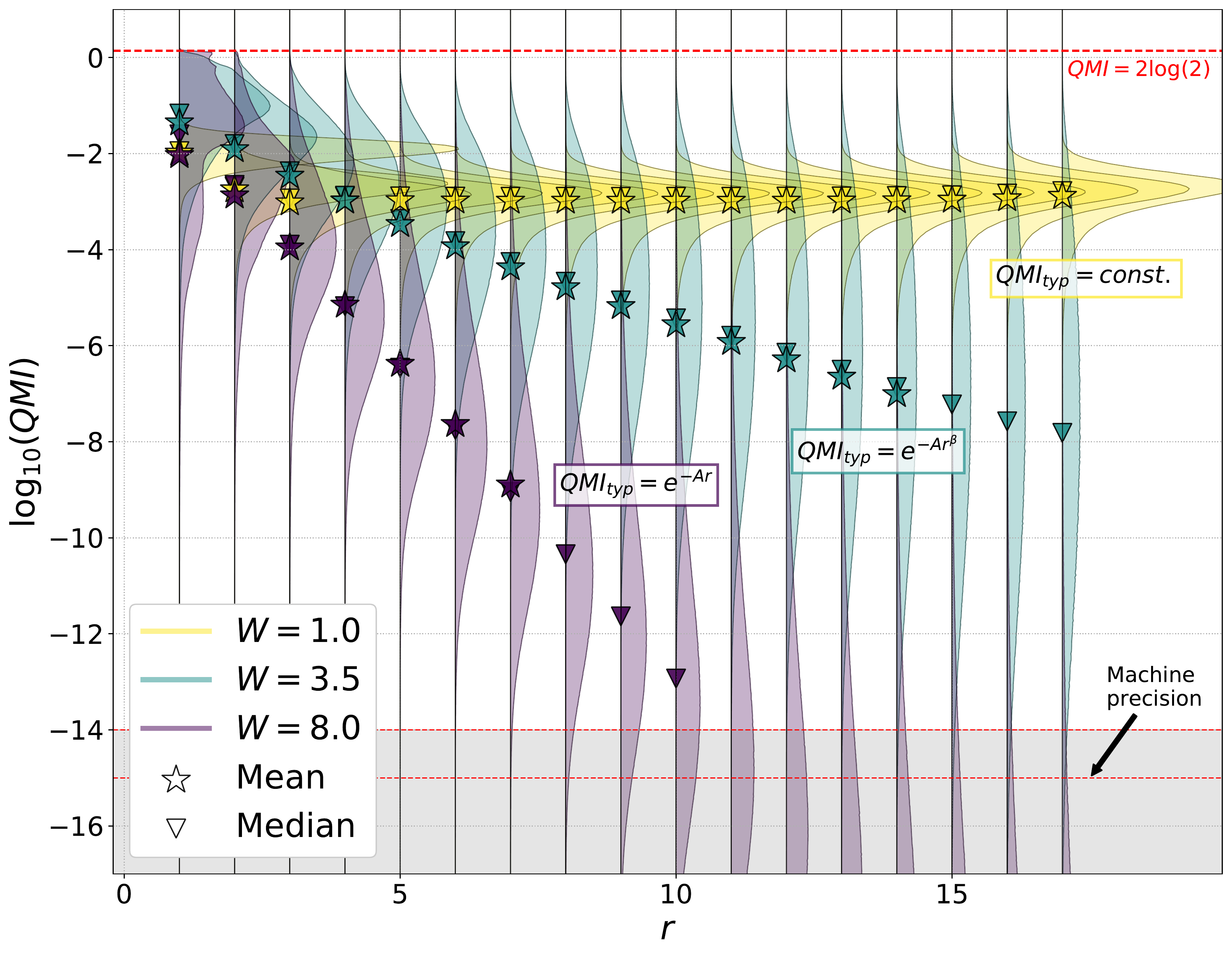}
\caption{\label{fig:histograms_qmi} Probability distributions of the logarithm of the two-site $QMI$ for a system of size $L = 18$ for all ranges $r = |i - j|$ (between sites $i$ and $j$) in the ergodic phase ($W = 1$), around the transition ($W = 3$), and deep in the MBL phase ($W = 8$).
Stars indicate the mean of the distributions and triangles indicate their median, which shows similar behavior.
While the typical (log-averaged) $QMI$ is constant with range in the ergodic phase, it decays exponentially deep in the MBL phase, and as a stretched exponential of the form $e^{-A r^\beta}$, with $1 / 2 < \beta < 1$, at moderate values the of disorder strength on the MBL side of the phase diagram.
At the transition, the decay follows a stretched exponential with $\beta= 1 / 2$, \emph{i.e.}, $QMI_{typ} = e^{-A \sqrt{r}}$.
We only consider distributions that are well above machine precision, \emph{i.e.}, those with at least 99\% of their mass above $QMI = 10^{-14}$ (Appendix~\ref{sec:app-precision}).
} 
\end{figure}

The eigenstates of the MBL and ergodic phases are qualitatively different in their properties, which affects the dynamical properties of their system.
MBL eigenstates show area-law entanglement~\cite{serbyn_local_2013,bauer_area_2013,yu_bimodal_2016,kjall_many-body_2014}, their correlations typically decay quickly with range~\cite{de_tomasi_quantum_2017}, and local observables vary wildly with energy, thus violating the eigenstate thermalization hypothesis (ETH)~\cite{peres_ergodicity_1984,deutsch_quantum_1991,srednicki_chaos_1994, srednicki_approach_1999,rigol_thermalization_2008,luitz_long_2016}.
On the contrary, ergodic eigenstates show volume-law entanglement, correlations typically decay slowly, and the ETH is satisfied as local observables vary smoothly with energy.
While much is known about eigenstates in the MBL and ergodic phases, the properties of eigenstates at the critical point between these phases are less well understood.
So far, the plurality of numerical evidence suggests localized eigenstates with sub-volume law entanglement~\cite{kjall_many-body_2014,luitz_many-body_2015,devakul_early_2015,lim_many-body_2016,khemani_critical_2017}, bimodality in the distribution of the entanglement entropy~\cite{yu_bimodal_2016}, and some forms of range-invariance~\cite{serbyn_criterion_2015,pekker_fixed_2017,gray_scale_2019,villalonga_eigenstates_2020}.
In addition, there is a body of work on renormalization group (RG) approaches to the phenomenology of the MBL transition~\cite{vosk_dynamical_2014,vosk_theory_2015,potter_universal_2015,zhang_many-body_2016,parameswaran_eigenstate_2017,thiery_microscopically_2017,goremykina_analytically_2019,dumitrescu_kosterlitz-thouless_2019,morningstar_renormalization-group_2019,morningstar_many-body_2020}.
Transitions in Floquet models have also been considered, with average long-distance correlations peaking at the transition showing system size independence~\cite{zhang_floquet_2016}.

In this work we focus on the correlations across a system throughout the MBL-ergodic phase diagram, providing extensive phenomenology in the MBL phase and at the transition.
Correlations have played key roles in understanding phases of matter and critical points.
In disordered systems, understanding the distribution of correlations, including their typical values, has been particularly insightful.
One canonical example of this is the random singlet phase, which appears as a universal fixed point in the strong disorder renormalization group (SDRG) analysis of many disordered ground state spin systems~\cite{fisher_random_1994,vidal_entanglement_2003,laflorencie_scaling_2005,vosk_many-body_2013,shu_properties_2016}.
In the random singlet phase, typical correlations exhibit stretched exponential behavior and universal features are anticipated for the full distribution of correlations~\cite{fisher_random_1994}. 

One way to quantify correlations is through the quantum mutual information (QMI).
The primary tool of this paper is the computation of the QMI in the spin-$1 \over 2$ nearest-neighbor antiferromagnetic Heisenberg chain with random onsite magnetic fields:
\begin{align}
\label{eq:model}
	H(W) = \frac{1}{4} \sum_{i=0}^{L-2} \vec{\sigma}_i \cdot \vec{\sigma}_{i+1} - \frac{W}{2} \sum_{i=0}^{L-1} h_i \sigma^z_i \text{.}
\end{align}
where the onsite magnetic fields $\left\{ h_i\right\}$ are sampled uniformly at random from $\left[ -1, 1 \right]$ and $W$ is the disorder strength.
The model of Eq.~\eqref{eq:model} has been studied extensively in the context of MBL~\cite{oganesyan_localization_2007,znidaric_many-body_2008,berkelbach_conductivity_2010,pal_many-body_2010,bauer_area_2013,luitz_many-body_2015,bar_lev_absence_2015,agarwal_anomalous_2015,bera_many-body_2015,luitz_extended_2016,luitz_long_2016,luitz_anomalous_2016,luitz_information_2017,yu_bimodal_2016,khemani_critical_2017,khemani_two_2017,serbyn_power-law_2016,de_tomasi_quantum_2017,villalonga_exploring_2018,herviou_multiscale_2019,gray_scale_2019,laflorencie_chain_2020}.
The two-site QMI was introduced in the context of MBL in Ref.~\cite{de_tomasi_quantum_2017}, where the authors found evidence for exponentially decaying $QMI$ with range in the MBL phase and slower decay in the ergodic phase.
Our goal will be to look at the distributions of the QMI considering both the typical and extreme (atypically strong) correlations.  

The key results of this work are the discovery of
\begin{itemize}
    \item  Stretched exponential behavior, $\exp\left[-Ar^\beta\right]$ (where $r$ is the range between two spins), of typical correlations both at the transition and in the MBL phase, spanning from $\beta=1/2$ at the transition and approaching $\beta=1$ around $W\approx 8$.
    Interestingly, the random singlet phase has the same decay of the typical correlations as the MBL-ergodic transition.
    \item Range-invariant universal (in the skewness and higher statistical moments) distributions of the $\log(QMI)$ at the transition.
    Even excess standard moments of these distributions are zero.
    \item Range-invariant \emph{strong} pairwise $QMI$ at the transition suggesting the existence of resonating cat states at all ranges at the critical disorder strength between the MBL and ergodic phases. 
\end{itemize}

Note the idea of a stretched exponential scaling of various quantities at moderate disorder has appeared in the MBL literature.
Refs.~\cite{zhang_many-body_2016,thiery_microscopically_2017,schiro_toy_2020} discuss a stretched exponential decay of the \emph{average} (not typical) correlations in the MBL phase from a simplified RG analysis, a microscopically motivated RG scheme, and a toy model, respectively. 
Refs.~\cite{dumitrescu_kosterlitz-thouless_2019,morningstar_renormalization-group_2019} consider instead the size of ergodic inclusions in the system; through RG arguments they find stretched exponential scaling of the size of these inclusions in the MBL phase and power law decay of their size at the transition.
Using a heuristic numerical algorithm, Ref.~\cite{herviou_multiscale_2019} finds evidence for the algebraic scaling of the cluster sizes at the transition, crossing over to a stretched exponential scaling of the cluster sizes upon entering the MBL phase and eventually becoming exponential at strong disorder strengths.

Our results are qualitatively different, finding strong numerical evidence for stretched exponential behavior of typical correlations both in the MBL phase \emph{and} at the transition.
Interestingly, our numerics are cleanest and most compelling at the critical point.
We note the stretched exponential behavior we find is clearly distinct from a power law.
It is an interesting open question how this compares to the average correlations found in various RG analyses and toy models as well as how typical correlations relate to the size of ergodic grains.
 
In Section~\ref{sec:typical} we analyze the structure of the typical correlations as well as look at the various moments of the $log(QMI)$.
In Section~\ref{sec:extreme} we show our results on the extremal values of the $QMI$ and their relation to scale invariant resonances.
In Section~\ref{sec:entanglement} we discuss the statistics of the second singular value of the bipartite entanglement entropy  which has been proposed recently as a robust order parameter in the ergodic-MBL phase diagram~\cite{samanta_extremal_2020}.
Finally, in Section~\ref{sec:conclusions} we summarize our findings and discuss their implications.

For $L = 18$, we obtain 100 eigenstates close to energy density $\epsilon \approx 0.5$ per disorder realization, over $10^4$ disorder realizations, obtaining a total of $10^6$ eigenstates.
For $L = 14, 16$, we obtain 5 eigenstates close to $\epsilon \approx 0.5$ per disorder realization, over a total of $2 \times 10^5$ disorder realizations, obtaining also a total of $10^6$ eigenstates per system size.
We do this for different values of the disorder strength $W$.

%% file: typical_correlations.tex
\section{Typical correlations}
\label{sec:typical}

In this section we look at the typical values of two-point correlations in an eigenstate of the Hamiltonian of Eq.~\eqref{eq:model} throughout the ergodic-MBL phase diagram.
We use the $QMI$ between all pairs of sites in a one-dimensional spin chain as a measure of the strength of their correlation that is agnostic to the choice of any particular correlation function.
The $QMI$ measures all correlations, both classical and quantum, between subregions in a system.
The $QMI$ between subregions $A$ and $B$ is defined as:
\begin{align}
\label{eq:qmi}
    QMI_{AB} \equiv S_A + S_B - S_{AB} \text{,}
\end{align}
where $S_A$ is the Von Neumann entanglement entropy between subsystem $A$ and its surroundings; we always work with the $QMI$ between pairs of sites, $i$ and $j$, which we denote $QMI_{ij}$.
The two-site $QMI$ has a maximum value of $2 \log(2)$, which occurs when two sites form a singlet.
However, in many-body systems it is very rare for two sites to form a singlet without being entangled to other sites; in the case of a multi-site singlet (i.e. linear superposition between two product states which differ on $k>2$ spins), the $QMI$ between two sites is equal to $\log(2)$.
We define $r$ as the range between two sites, \emph{i.e.}, $r \equiv |i - j|$.
Ref.~\cite{de_tomasi_quantum_2017} finds that the typical values of the $QMI$ decay exponentially with $r$ in the MBL phase and slower than exponentially in the ergodic phase.
Here we focus in detail on the question of the behavior of the typical correlations along a one-dimensional system in the ergodic-MBL phase diagram.

We work with the distributions of the $\log(QMI)$ (see Fig.~\ref{fig:histograms_qmi}, where, for readability, the $\log_{10}(QMI)$ is presented), as opposed to the distributions of the $QMI$.
We consider the $\log(QMI)$ for each range $r$ separately.
A first visual inspection shows compact distributions that are constant across ranges at weak disorder and decaying and broadening (with $r$) distributions at moderate and large disorder.
Also, the distributions seem skewed in opposite directions at large and small disorder strengths.

In Section~\ref{sec:means} we study the decay of the typical correlations with $r$; surprisingly, we find a region in the MBL side of the phase diagram with a stretched exponential decay at moderate values of the disorder strength $W$ terminating at the transition with a stretched exponential with exponent $1 / 2$; this has similarities with the random singlet phase that arises as a fixed point in renormalization group studies of disordered systems~\cite{fisher_random_1994,vosk_many-body_2013,shu_properties_2016}.
In Section~\ref{sec:std} we look at the standard deviation of these distributions, finding they increase linearly with range $r$.
Next, in Section~\ref{sec:skewness}, we study the skewness and higher statistical moments of the distributions; our results show that these moments take a universal value at the transition for large enough ranges.
This implies that the distribution of $\log(QMI)$ is universal at the transition beyond the first two moments. 
Finally, in Section~\ref{sec:together} we summarize and discuss our findings on the typical correlations.
As we can see in Fig.~\ref{fig:histograms_qmi}, the QMI reaches machine precision ($\approx 10^{-15}$) at large range $r$ and large disorder strength $W$; we only consider those points (\emph{i.e.} the triplet $(L, W, r)$) for which the distribution of the $\log(QMI)$ has at least 99\% of its mass above $QMI = 10^{-14}$, \emph{i.e.}, one order of magnitude above the machine precision threshold of double-precision floating-point numbers (see Appendix~\ref{sec:app-precision}).

\subsection{The decay of $\text{QMI}_{typ}$}
\label{sec:means}

\begin{figure}[t]
\centering
\includegraphics[width=1.00\columnwidth]{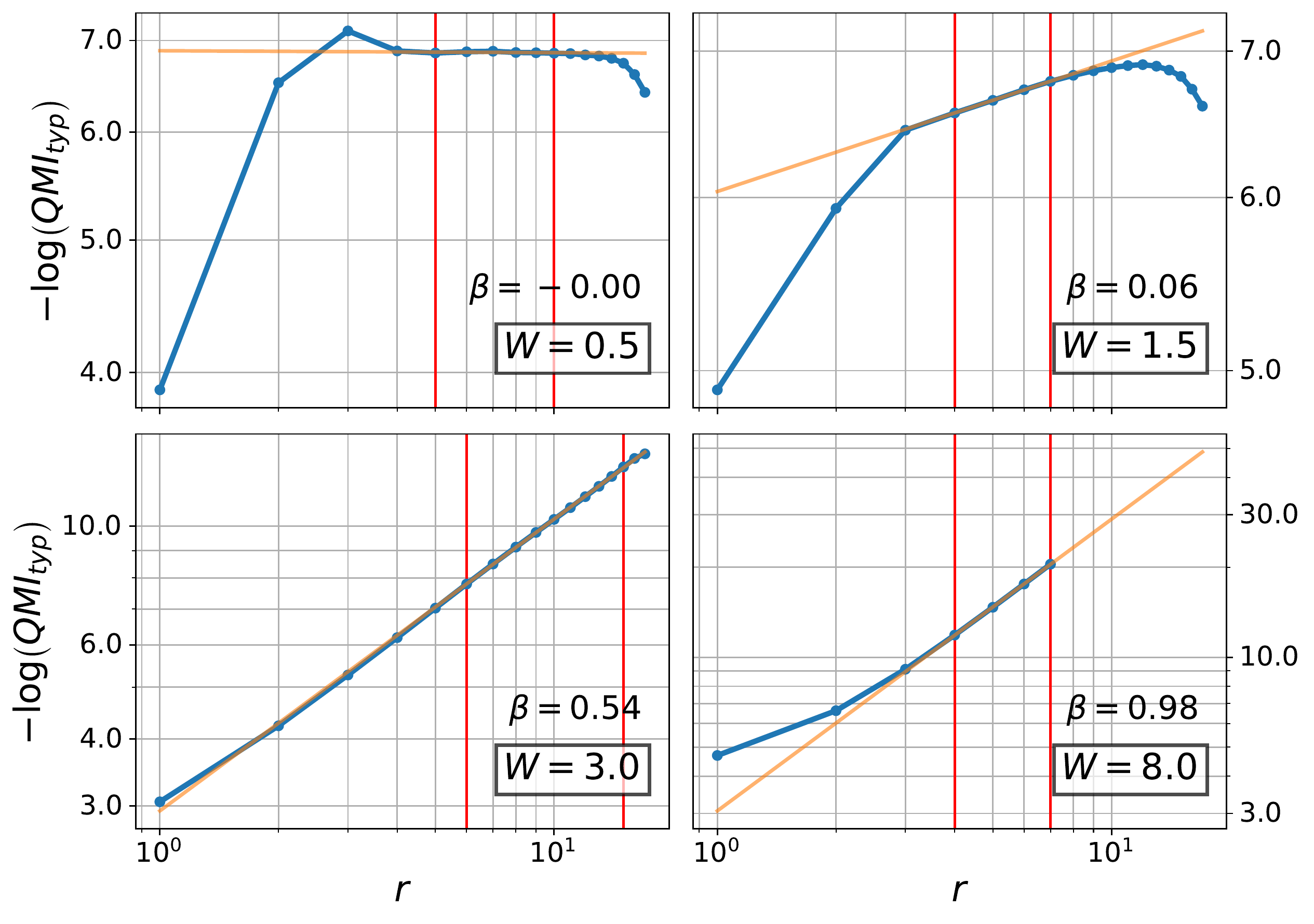}
\caption{\label{fig:means_examples} Log-log plot of $-\log\left(QMI_{typ}\right)$ as a function of $r$ for a system of size $L = 18$.
We can see that for moderate and large disorder strength ($W = 3.0$ and $8.0$ in the plots) the stretched exponential ansatz ($QMI_{typ} = e^{-A r^\beta}$) fits well the data at large $r$.
A linear fit to the curves is shown, as well as the interval of data taken for the fit (red vertical lines).
Deep in the ergodic phase ($W = 0.5$), $QMI_{typ}$ is constant.
At slightly higher values of the disorder strength ($W = 1.5$), it is unclear what the functional form of the curve is, since the fit to a stretched exponential is not reliable.
See Appendix~\ref{sec:app-beta_fits} for fits over all values of $L$ and $W$.
}
\end{figure}

The typical values of the $QMI$ are defined as the log-averaged $QMI$:
\begin{align}
\label{eq:typical_qmi}
  QMI_{\text{typ}} \equiv \left\langle QMI \right\rangle_{\text{log}} = e^{\left\langle \log(QMI) \right\rangle} \text{,}
\end{align}
\emph{i.e.}, it is computed by exponentiating the mean of the distributions of Fig.~\ref{fig:histograms_qmi}.
We find that $QMI_\text{typ}$ fits a stretched exponential of the form
\begin{align}
\label{eq:ansatz}
    QMI_{\text{typ}} = e^{-A r^\beta},
\end{align}
at large range $r$ in the MBL phase and at the transition.
This is demonstrated by the linear behavior on the log-log plot of $-\log\left( QMI_{typ} \right)$ in Fig.~\ref{fig:means_examples}.
This linear fit is especially compelling at the transition ($W \approx 3.0$ for $L = 18$) where essentially all ranges are well fit by a linear curve.
In the ergodic phase, at intermediate values of the disorder strength ($W = 1.5$) the linear fit is of poor quality, and deep in the ergodic phase ($W = 0.5$) we find that $QMI_{typ}$ is constant with $r$.

\begin{figure}[t]
\centering
\includegraphics[width=1.00\columnwidth]{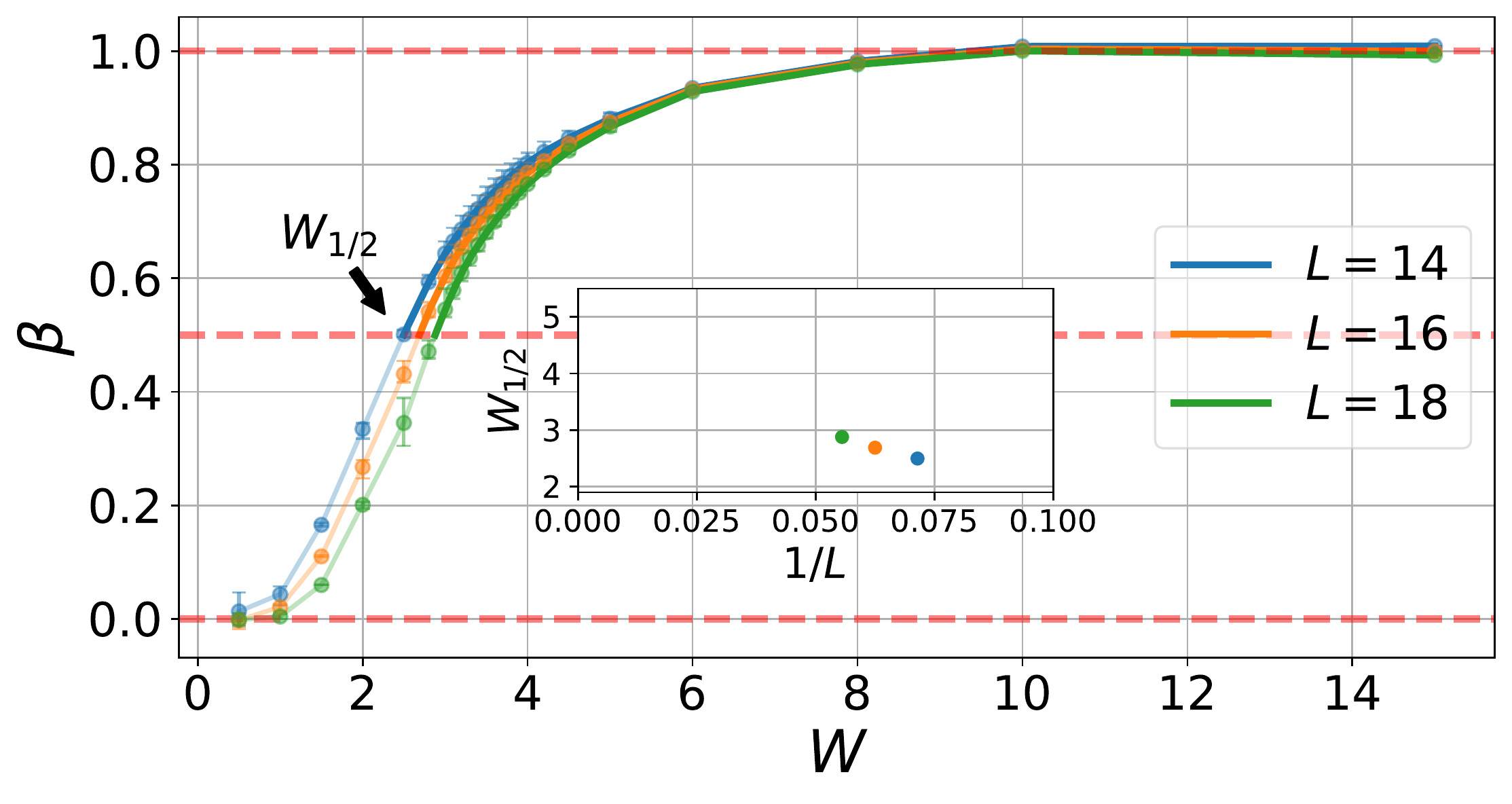}
\includegraphics[width=1.00\columnwidth]{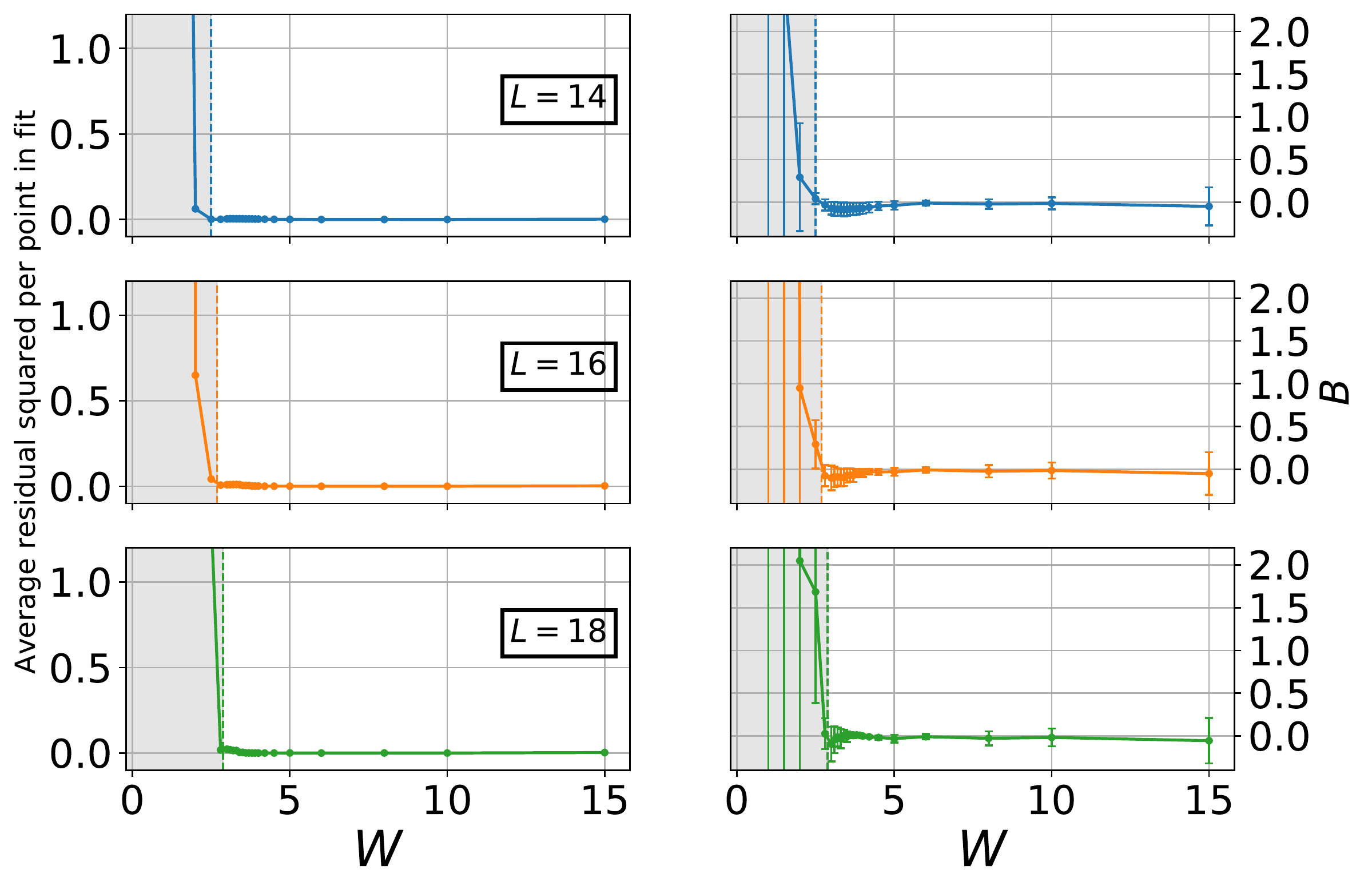}
\includegraphics[width=1.00\columnwidth]{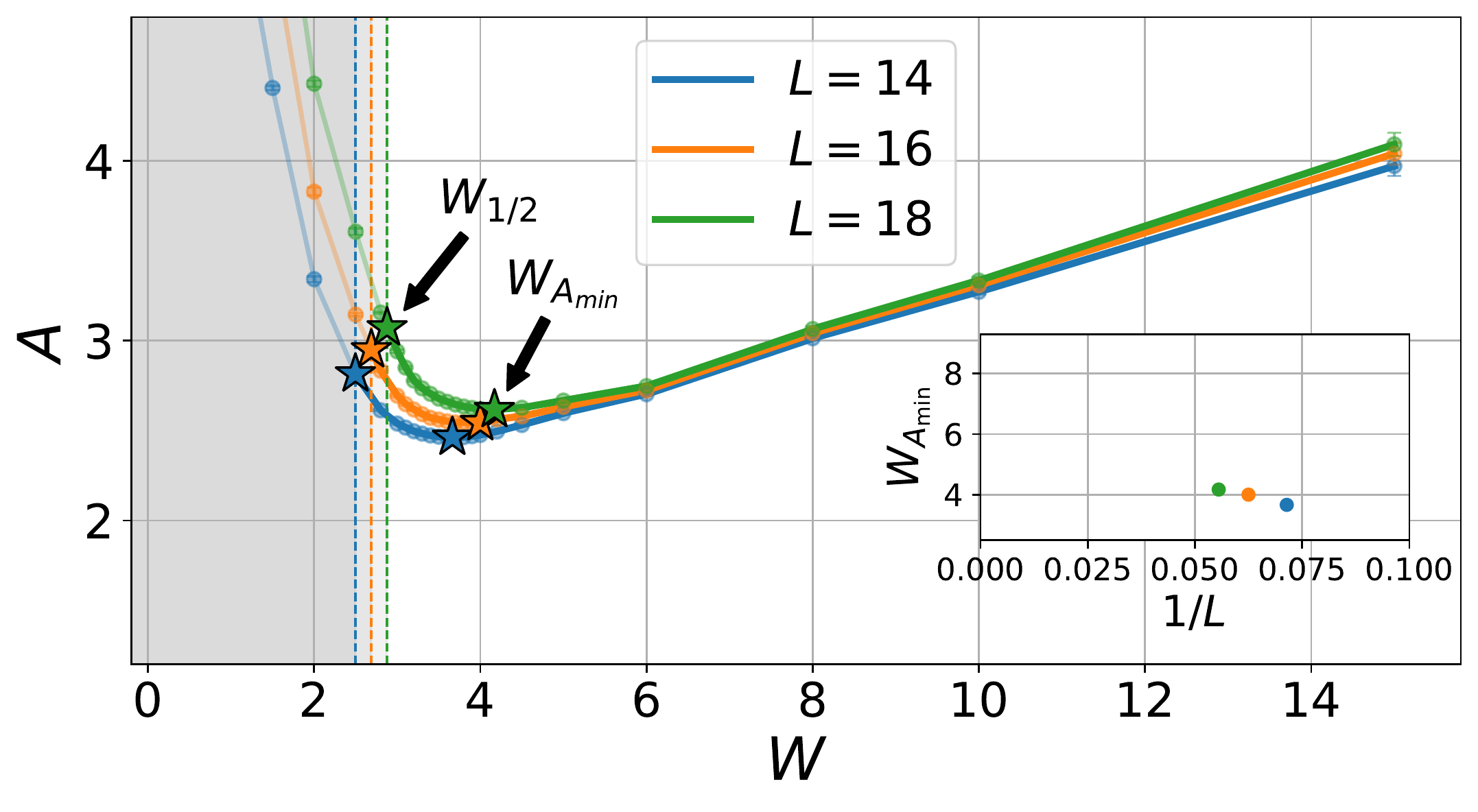}
\caption{\label{fig:beta} 
\textbf{Top:} Exponent $\beta$ of the stretched exponential of the decay of the typical $QMI$, $QMI_{typ} = e^{-A r^{\beta}}$.
Inset shows $W_{1/2}$ (at which $\beta = 1/2$) as a function of $1/L$ ($W_{1/2} = 2.50, 2.69, 2.88$ at $L = 14, 16, 18$, respectively).
\textbf{Middle left:} Average residual squared per point in the linear fit $QMI_{typ}^{1 / \beta} = -A r + B$.
\textbf{Middle right:} $B$ as a function of $W$.
$B$ is zero if the stretched exponential assumption was correct ($W > W_{1 / 2}$).
\textbf{Bottom:} coefficient $A$ as a function of $W$.
Inset shows $W_{A_\textrm{min}}$ as a function of $1/L.$
\textbf{Middle and bottom:} We show dashed vertical lines at the values of $W_{1/2}$ extracted from the curves in the top panel.
} 
\end{figure}

We can extract the exponent $\beta$ from the slope of the fit to the log-log plot.
The values of $\beta$ are presented in the top panel of Fig.~\ref{fig:beta} (confidence intervals are defined by the maximum (minimum) $\beta$ found over all linear fits of three or more consecutive points in the region fitted; see Appendix~\ref{sec:app-beta_fits}).
As discussed above, the values of $\beta$ are not reliable at low disorder strength; despite this, we present all values of $\beta$ even when not reliable.
By visual inspection, we consistently find, across different system sizes $L$, that the fits from which we extract $\beta$ are of good quality above $W_{1 / 2}$, which we define as the value of $W$ at which $\beta = 1 / 2.$
Interestingly, the data on the log-log plots from which $\beta$ is extracted falls below the linear fit at low ranges for weak disorder, while it lays above the linear fit at large values of $W$.
All ranges, including low range data, fall exactly on top of the linear fit precisely when $\beta = 1 / 2$ (see Appendix~\ref{sec:app-beta_fits} for additional data).

The stretched exponential behavior therefore seems to extend from $W_{1 / 2}$ up to a value of $W$ for which $\beta \approx 1$.
The inset of Fig.~\ref{fig:beta} shows $W_{1 / 2}$ as a function of $1 / L$ for the three values of $L$ we measure.
A naive extrapolation to $L \rightarrow \infty$ seems consistent with $W_{1 / 2}$ coinciding with the critical value of $W$ in the thermodynamic limit, \emph{i.e.}, $W_{1/2} (L \rightarrow \infty) = W_c \approx 4$.

In order to back our observation that the decay of $QMI_{typ}$ follows a stretched exponential (Eq.~\ref{eq:ansatz}) down to the value of $W$ for which $\beta = 1 / 2$ ($W_{1 / 2}$), we present in the middle-left panel of Fig.~\ref{fig:beta} the average residual squared per point in the fits from which $\beta$ was extracted, \emph{i.e.}, log-log plots like those of Fig.~\ref{fig:means_examples}.
We can see that the residuals are consistent with high-quality fits at and above $W_{1 / 2}$, where they are practically zero.
Below $W_{1 / 2}$ (shaded our region) the residuals per point rapidly increase.

Finally, the bottom panel of Fig.~\ref{fig:beta} shows the values of $A$ in the stretched exponential as a function of $W$ for different system sizes $L$.
We extract $A$ from the slope of a linear fit of $\log(QMI_{typ}) ^ {1 / \beta} = A^\prime r + B$, where $\beta$ takes the empirically obtained value of the top panel of Fig.~\ref{fig:beta}, and $A^\prime = A^{1 / \beta}$.
Note that this is only correct if $B = 0$ in the fit, which simultaneously corresponds to our stretched exponential ansatz being correct.
Indeed, these fits find $B$ to be practically zero (within error bars) for $W \geq W_{1 / 2}$, as shown in the middle-right panel of Fig.~\ref{fig:beta}, which is an excellent \emph{a posteriori} consistency check for our ansatz, independent of the residuals of the middle-left panel.
On the contrary, $B$ grows rapidly below $W_{1 / 2}$, where the ansatz breaks.
As in the case of $\beta$, we show in the bottom panel all values of $A$ found, regardless of their reliability.
We have highlighted two sets of points (marked as stars).
First, the values of $A(W_{1 / 2})$ show an increasing trend as $W_{1 / 2}$ shift towards higher values of $W$ with system size; we will revisit this in Section~\ref{sec:together}.
Second, we drive the reader's attention to the points at which $A$ is minimal, $W_{A_{min}}$.
The $QMI$ decays exponentially deep in the MBL phase, \emph{i.e.}, $QMI_{typ} = e^{-A r}$; since the system should localize further as $W$ increases, $A$ must increase with $W$ if the decay is exponential.
Note $1 / A$ is a localization length.  
For this reason, we anticipate that the decay must be a stretched exponential out to at least $W_{A_{min}}$, which we regard as a lower bound for the value of $W$ at which the decay transitions from stretched exponential to exponential: $W_1$.
Our data suggests that that $W_{1 / 2} < W_1$ and $W_1 < \infty$ in the thermodynamic limit, a situation where the stretched exponential decay region is stable over a region in the MBL phase before it becomes an exponential decay.
However, we cannot rule out two other scenarios in which either $W_{1 / 2} \rightarrow W_1$ in the thermodynamic limit or $W_{1} \rightarrow \infty$.

\subsection{The standard deviation}
\label{sec:std}

\begin{figure}[t]
\centering
\includegraphics[width=1.00\columnwidth]{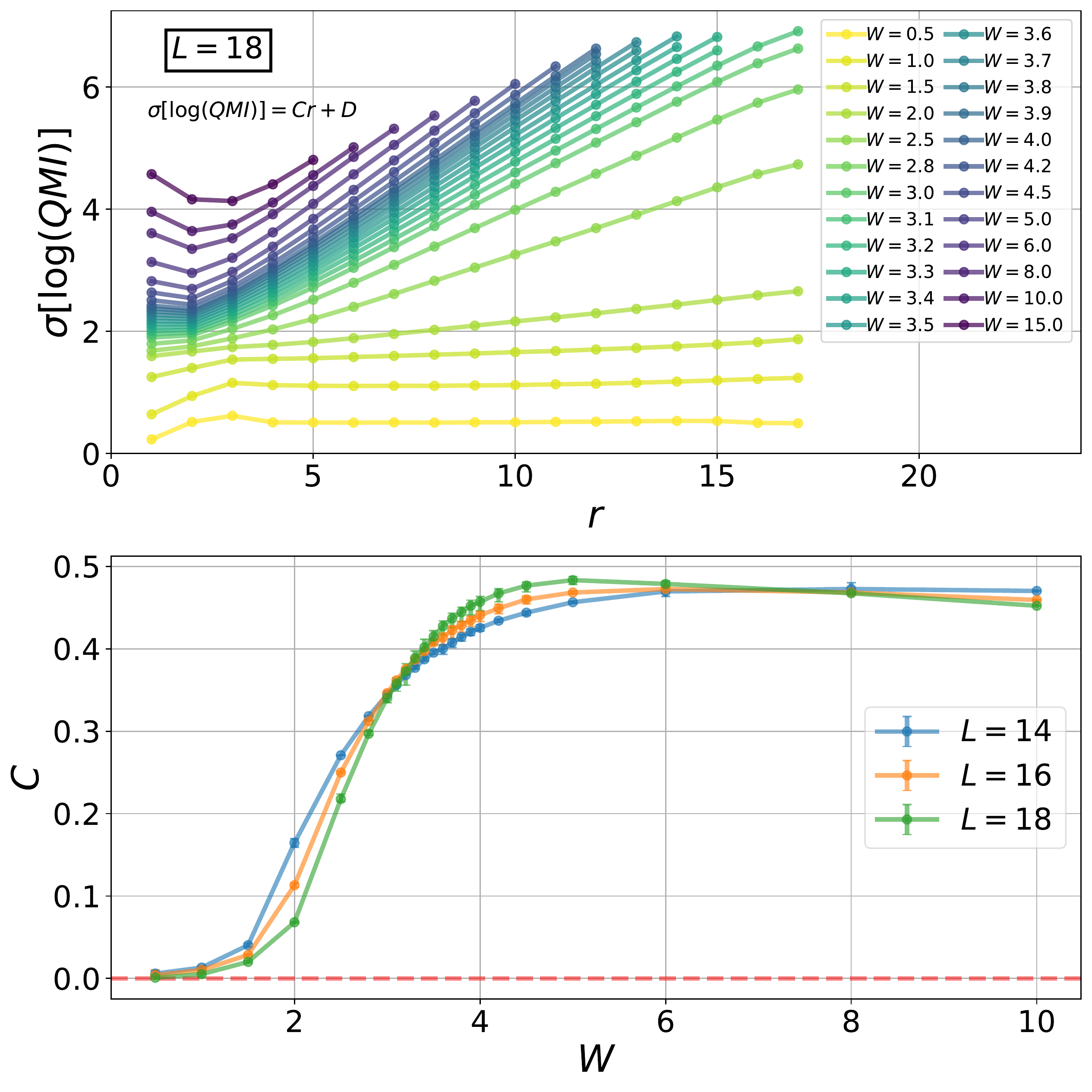}
\caption{\label{fig:stds} \textbf{Top:} Standard deviation of $\log(QMI)$ as a function of range $r$ for different values of the disorder strength $W$, for systems of size $L = 18$.
We can see at sufficiently large $r$ that the scaling is linear with $r$.
Linear fits at $1 \lessapprox W \lessapprox 2$ and $W \gtrapprox 8$ are of lower quality than other values of $W$.
Note that points affected by the finite machine precision have been removed and that finite size effects are present at the largest values of $r$.
\textbf{Bottom:} Slope $C$ of the linear fit of $\sigma \left[ \log(QMI) \right] = C \cdot r + D$ for different system sizes.
The transition region from $C \approx 0$ to $C \approx 0.5$ seems to be consistent with
typical estimates of the critical disorder strength.  
At large $W$ our data shows $C$ slowly dropping with $W$; it is unclear whether this is affected by the lack of large $r$ points at large $W$.
At $W = 15$ we do not have large enough ranges to reach the linear scaling regime.
Confidence intervals are defined by the maximum (minimum) value of $C$ found over all linear fits of three or more consecutive points in the region fitted (see Appendix~\ref{sec:app-C_fits}).}
\end{figure}

We use the standard deviation of the distributions of $\log(QMI)$ as a measure of their width.
It is already apparent from Fig.~\ref{fig:histograms_qmi} that the width of the distributions deep in the ergodic phase is constant.
Around the transition and in the MBL phase, the width increases with $r$.

We present $\sigma\left[ \log(QMI) \right]$ as a function of $r$ in the top panel of Fig.~\ref{fig:stds}.
These curves (after eliminating distributions affected by machine precision, and ignoring finite size effects at large range $r$) follow linear scaling as a function of $r$ of the form $\sigma\left[ \log(QMI) \right] = C r + D$.
The lower panel of Fig.~\ref{fig:stds} shows the values of $C$ as a function of $W$ for different system sizes.
Constant $\sigma\left[ \log(QMI) \right]$ deep in the ergodic phase gives $C = 0$.
In the MBL phase $C \lessapprox 0.5$, dropping slowly (or staying nearly constant) as $W$ increases. 
Between these two extremes at small and large $W$, there is a rapid increase in $C$ from $C \approx 0$ to $C \approx 0.5$, which gets sharper at larger system sizes $L$.
The curves at different $L$ cross at a value of $W$ which is within the range of typically estimated values of the critical $W_c$ and we can treat this as a poor man's scaling collapse (our attempts of carrying out a more formal scaling collapse were unsuccessful at generating reliable results).

\subsection{The skewness and higher moments}
\label{sec:skewness}

\begin{figure}[t]
\centering
\includegraphics[width=1.00\columnwidth]{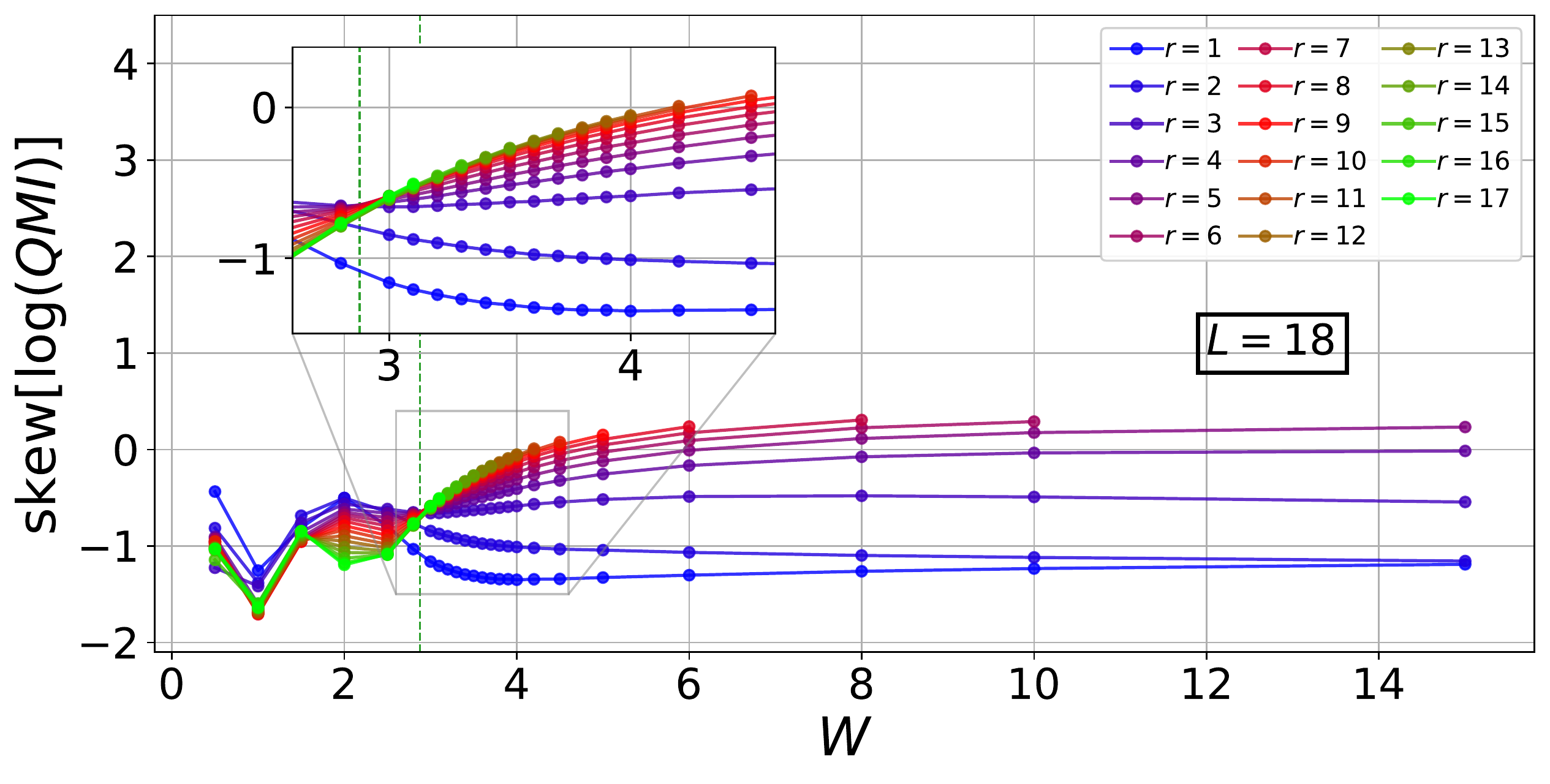}
\includegraphics[width=1.00\columnwidth]{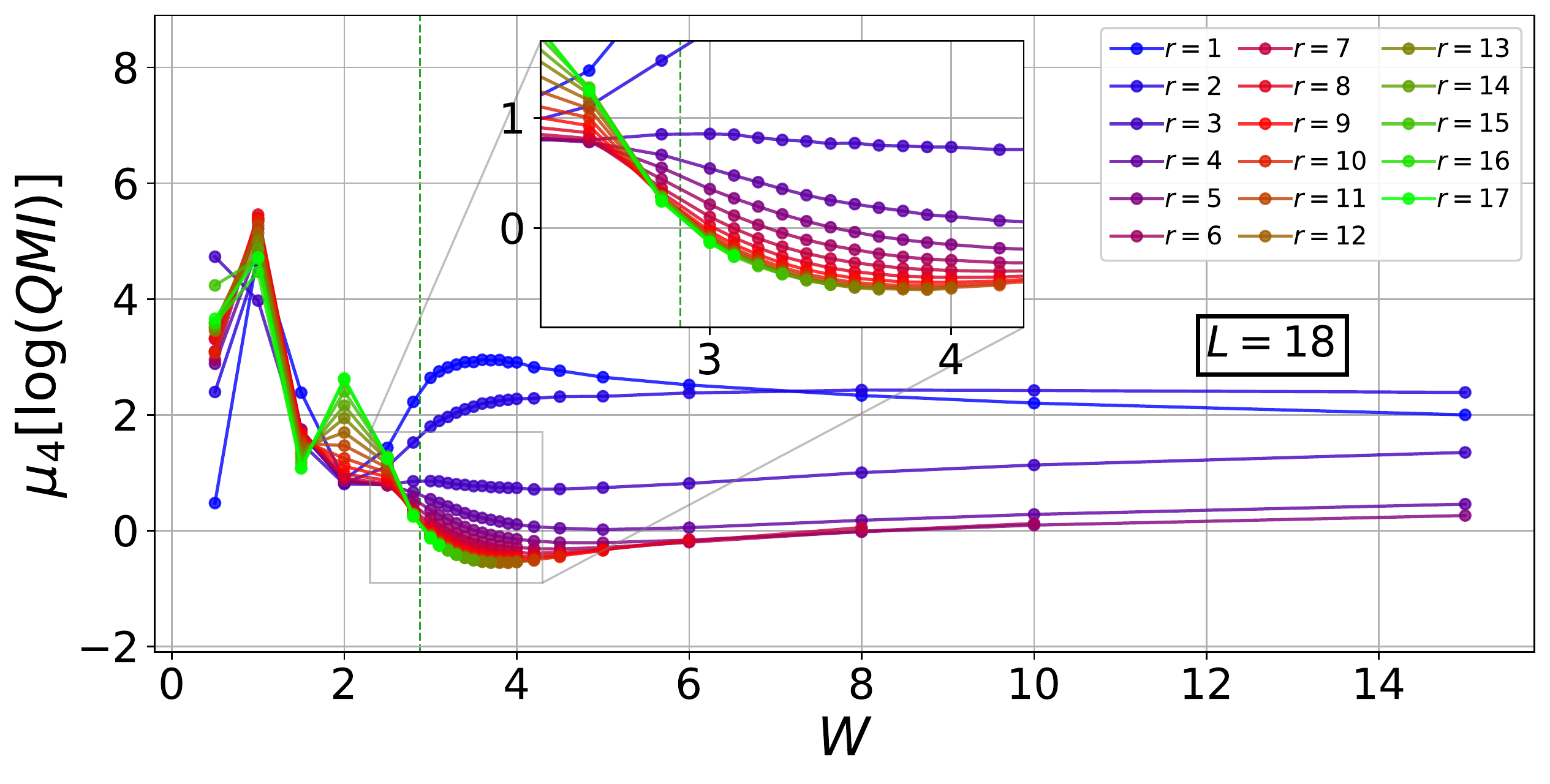}
\caption{\label{fig:skewness}
Third excess moment (skewness; $\mu_3$) and fourth excess moment ($\mu_4$) of $\log(QMI)$ as a function of $W$ for all ranges $r$ for a system of size $L = 18$.
Vertical dashed lines at $W = W_{1/2}$ (where we independently find $\beta = 1/2$ in Section~\ref{sec:means}) are shown.
In both cases, the moments are range invariant close to $W_{1 / 2}$.
Distributions are positively skewed at large $W$ (for large enough $r$) and negatively skewed at weak $W$.
At the range invariant point the skewness is close to -0.65.
At the range invariant point $\mu_4 \approx 0$.
Higher odd (even )moments show similar behavior as the skewness ($\mu_4$) (see Fig.~\ref{fig:moments_colormaps}).
} 
\end{figure}

\begin{figure*}[t]
\centering
\includegraphics[width=2.00\columnwidth]{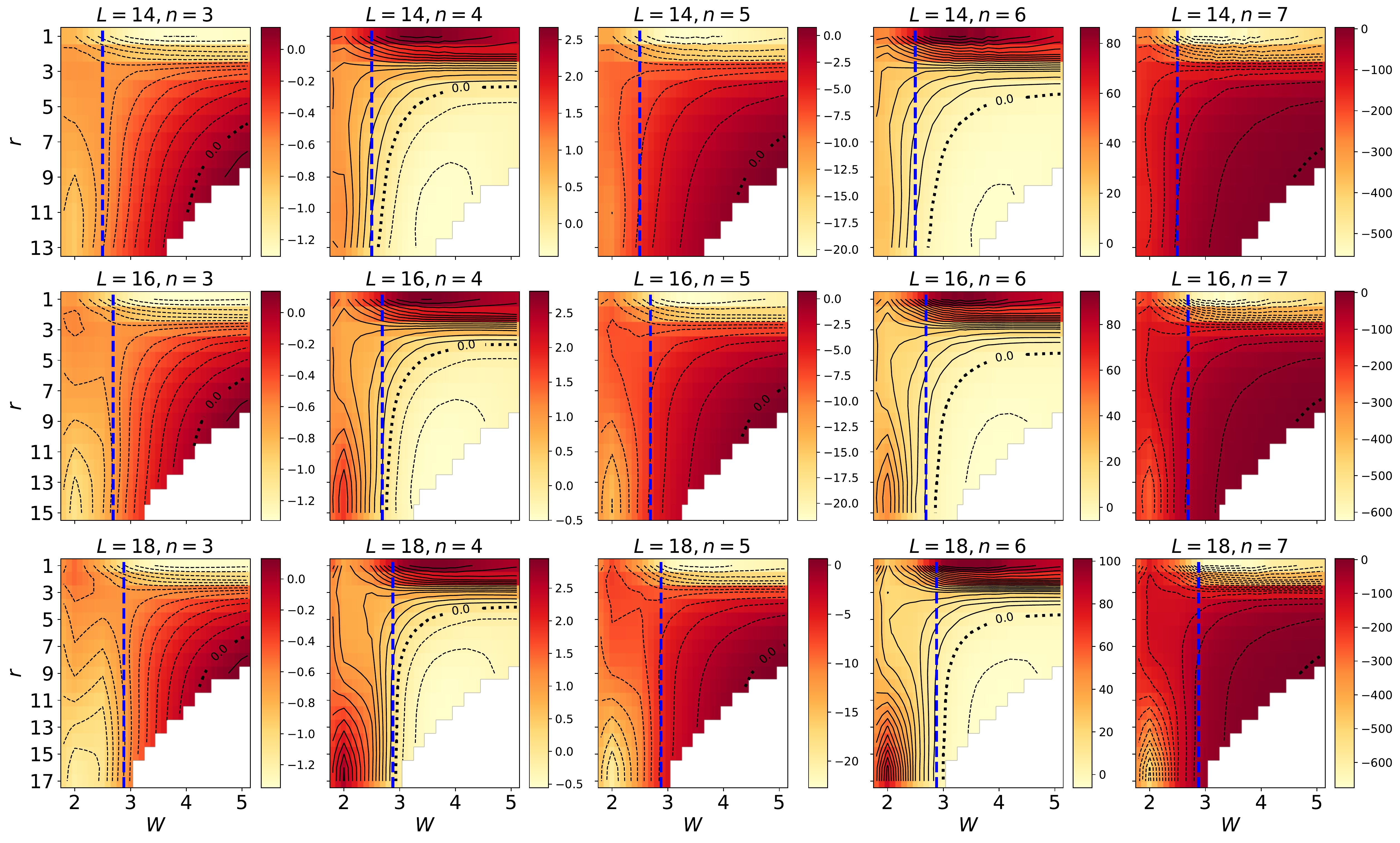}
\caption{\label{fig:moments_colormaps}
Colormap of the excess standardized moment $\mu_n$ for $n = 3, 4, 5, 6, 7$ and different system sizes $L$; $\mu_n$ is linearly interpolated across $W$ and contour lines are added for clarity; as usual, large $W$ and $r$ distributions that were affected by the finite machine precision are removed.
Vertical, dashed blue lines show the value of $W_{1/2}$ at which the typical $QMI$ decays as a stretched exponential with $\beta = 1/2$.
Contour lines show $\mu_n$ becomes range invariant close to $W_{1/2}$.
This agreement is very close at smaller values of $n$; at larger values of $n$, however, we see the agreement becoming closer as $L$ grows.
In addition, our results suggest even moments being range invariant at a value of zero (dotted contour line).
}
\end{figure*}

Fig.~\ref{fig:histograms_qmi} shows that the distributions of $\log(QMI)$ are skewed negatively deep in the ergodic phase, and positively deep in the MBL phase (at large enough $r$), with perhaps a more symmetric, close to unskewed shape around the transition.
In this section we study the skewness of these distributions, as well as their higher-order statistical moments.

The excess moment of order $n$ of a distribution over a random variable $x$ is defined as:
\begin{align}
\label{eq:moment}
    \mu_n \equiv \frac{\mathbb{E}\left[ (x - \langle x \rangle)^n \right]}{\sigma^n} - \mu_n^{\textrm{norm.}} = \mu_n^{\textrm{stand.}} - \mu_n^{\textrm{norm.}}\text{,}
\end{align}
where $\mu_n^{\textrm{stand.}}$ is the standardized moment of order $n$ (normalized by the the $n$'th power of the standard deviation) and $\mu_n^{\textrm{norm.}}$ is the $n$'th moment of the normal distribution, which by definition has all excess moments equal to zero.
$\mu_n^{\textrm{norm.}}$ is zero for odd $n$ and $\sigma^n (n - 1)!!$ for even $n$. 
The third standardized moment is called the \emph{skewness}.

In Fig.~\ref{fig:skewness} we present the skewness ($\mu_3$) and $\mu_4$ of $\log(QMI)$ as a function of $W$ for each range $r$ for a system of size $L = 18$.
As expected, the skewness is negative at small $W$ and positive for large ranges $r$ at large values of $W$.
Interestingly, as seen more clearly in the inset of Fig.~\ref{fig:skewness}, the curves of different ranges cross at a point that is close to $W_{1 / 2}$ for $L = 18$ (vertical dashed line), which was estimated independently in Section~\ref{sec:means} as the disorder strength at which $\beta = 1 / 2$.
$\mu_4$ also becomes range invariant close to $W_{1 / 2}$.
In addition, the scale invariant value of $\mu_4$ is close to zero.

We now proceed to inspect the excess moments in a more systematic way.
Fig.~\ref{fig:moments_colormaps} shows colormaps of $\mu_n$ as a function of $W$ and $r$.
We see that odd (even) moments look alike.
In all cases the moments become range invariant (at large enough ranges) close to $W_{1 / 2}$ (independently computed in Section~\ref{sec:means}) with qualitatively different behavior between larger and smaller $W$.
The difference between the apparent range-invariant value of $W$ and $W_{1 / 2}$ decreases quickly with $L$;  for odd moments, even at small $L$,  $W_{1/2}$ is already very close to the range-invariant value of $W$.  Interestingly, for even moments (but not odd moments), the $\mu_n = 0$ contour line is essentially at $W_{1/2}$ at large enough $r$ and $L=18$.
Finally, we note that the ranges at which $\mu_n$ shows range invariant behavior become larger with $n$; in addition, in all cases we observe slight finite size effects at the largest ranges.

\subsection{Putting it all together}
\label{sec:together}

In summary, the typical correlations in a one-dimensional spin chain of the model in Eq.~\eqref{eq:model} decay exponentially deep in MBL.
Deep in the ergodic region, correlations are constant with range $r$. 
At moderate disorder strength, and above $W_c$, ($W_c \leq W \leq W_1$), typical correlations decay as a stretched exponential ($QMI_{typ} = e^{-A r^\beta}$), which takes the form $QMI_{typ} = e^{-A \sqrt{r}}$ at the transition (\emph{i.e.}, $W_{1 / 2} = W_c$) and the exponential form ($\beta = 1$) at $W_1$.
Our results suggest both this stretched exponential and exponential decay region of the phase diagram are stable in the thermodynamic limit
\footnote{
We can however not rule out scenarios in which: (1) $W_1 \rightarrow \infty$ in the thermodynamic limit, (2) $\beta > 1$ (compressed exponential decay) for $W > W_1$, (3) $W_{1 / 2} \rightarrow W_1$ in the thermodynamic limit, (4) $W_{1 / 2} \neq W_c$, or compatible combinations of them.
}

The distributions of $\log(QMI)$ have constant spread (standard deviation) deep in the ergodic phase.
At moderate and strong disorder strengths, they broaden linearly with range $r$.

Our results show various similarities between the ergodic-MBL transition and the random singlet phase, which emerges as an infinite disorder fixed point in strong disorder renormalization group studies of the ground states of disordered spin systems.
Typical correlations, which decay as a stretched exponential with $\beta=1/2$ are found in the random singlet phase.
In addition, it is anticipated~\cite{fisher_random_1994} that the random singlet phase has invariance of the distributions of the logarithm of the correlations divided by $\sqrt{r}$.
Our results are consistent with this for all standardized moments (up to the 7'th);
note however the standard deviation (and also the variance, \emph{i.e.}, the second moment), does not collapse even under the $\sqrt{r}$ rescaling.
This might be regarded as the ergodic-MBL transition satisfying a weaker version of universality as conjectured in Ref.~\cite{fisher_random_1994} for the random singlet phase.
While we find zero even excess moments, odd moments appear to converge to non-zero values.

There is a paradox in the fact that at $W_{1 / 2}$ the distribution of $\log(QMI)$ takes a universal form with a mean that decays with $\sqrt{r}$, while its standard deviation increases as $C r + D$.
Such family of distributions would quickly (as $r$ increases) have half of their weight above $QMI_{max} = 2 \log(2)$, which is an upper bound for the $QMI$.
In order for these scalings (mean and standard deviation) to be compatible with a fixed distribution of $\log(QMI)$ at long range, the area under the distribution that lays above $QMI_{max}$ has to vanish with $r$, or at least stay constant.
The only way out of this paradox is a coefficient $A(W = W_{1 / 2}, L)$ that increases at least as fast as $L^{1 / 2}$ with system size, but not with a smaller exponent.
This way, larger values of $r$ are only encountered for large values of $L$, which guarantee a large enough coefficient $A$, and thus enough room for the distribution to broaden while staying mostly below the $2 \log(2)$ threshold.
Our results (see lower panel of Fig.~\ref{fig:beta}, $W_{1 / 2}$ stars) are compatible with this scaling; however, given the small amount of data (only three small values of $L$), we cannot make any reliable claim.
In general, in the stretched exponential decay region, we require $A(W, L)$ to scale at least as $L^{1 - \beta}$.

%% file: extreme_correlations.tex
\section{Extreme correlations}
\label{sec:extreme}

\begin{figure}[t]
\centering
\includegraphics[width=1.00\columnwidth]{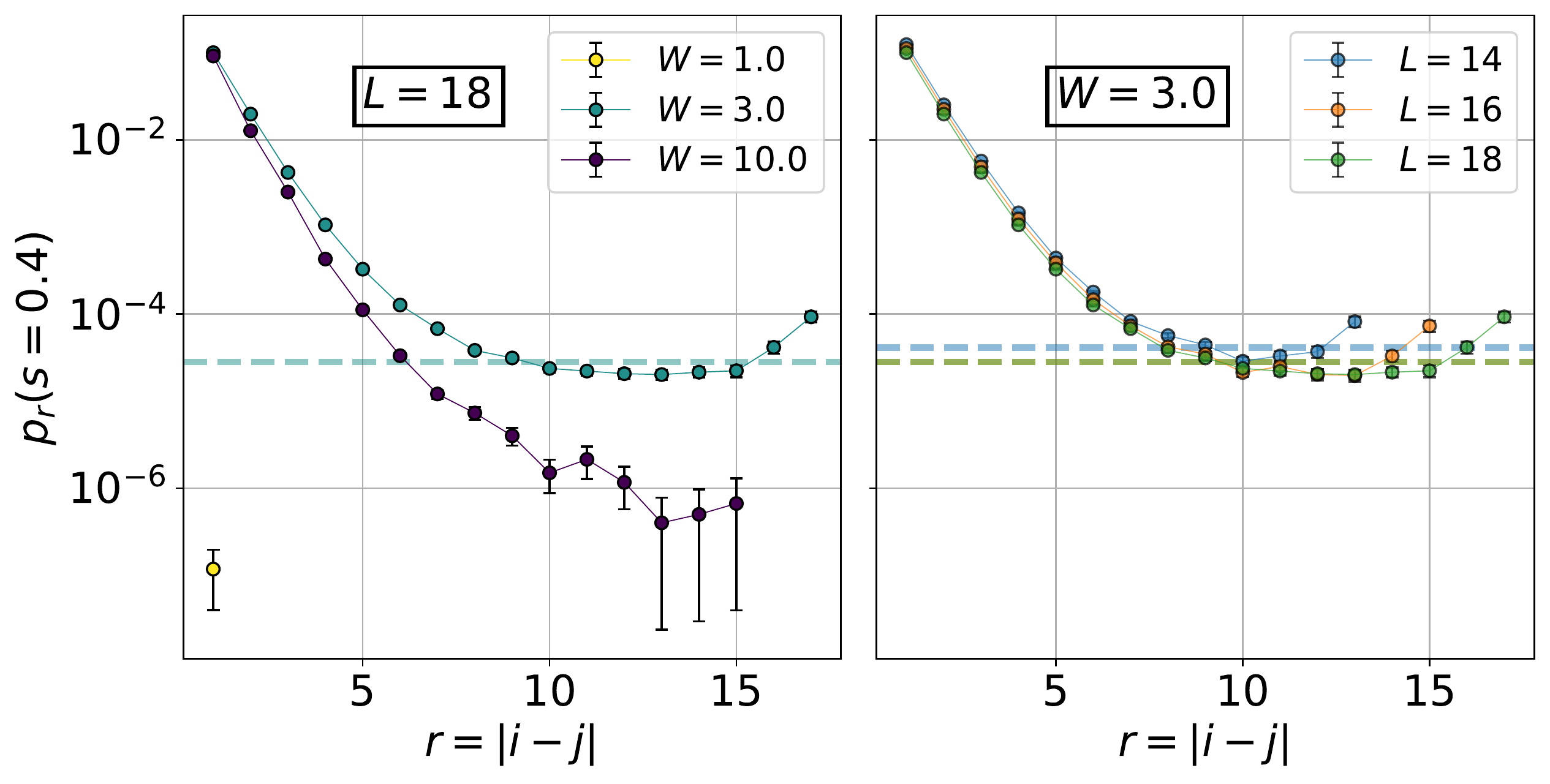}
\caption{\label{fig:prob_qmi} Probability $p_r(s=0.4)$ (see Section~\ref{sec:extreme}) given two sites $i$ and $j$, with $r=|i-j|$, as a function of $r$ for different disorder strengths $W$ (left) and $L$ (right).
Unreported data points correspond to values for which we had no samples and so were unable to estimate $p_r(s)$.
Deep in the MBL phase ($W = 10$), $p_r(s)$ decays with $r$, in line with the localization of correlations.
Deep in the ergodic phase ($W = 1$), correlations are small and have small spread in their order of magnitude (see Section~\ref{sec:typical}) so finding a value of the $QMI$ that exceeds $s$ is improbable.
Around the transition ($W = 3$), the probability of finding strong two-site $QMI$ bonds becomes range invariant, \emph{i.e.}, is constant as a function of $r$ at sufficiently large $r$ (and away from finite size effects at very large $r$).
Errors represent the standard deviations of the distribution of $p_r(s)$  over 200 bootstrapping resamples of the disorder realizations.
Dotted lines represent the saturated value of $p_r(s)$, $p_{sat}$.
} 
\end{figure}

\begin{figure}[t]
\centering
\includegraphics[width=1.00\columnwidth]{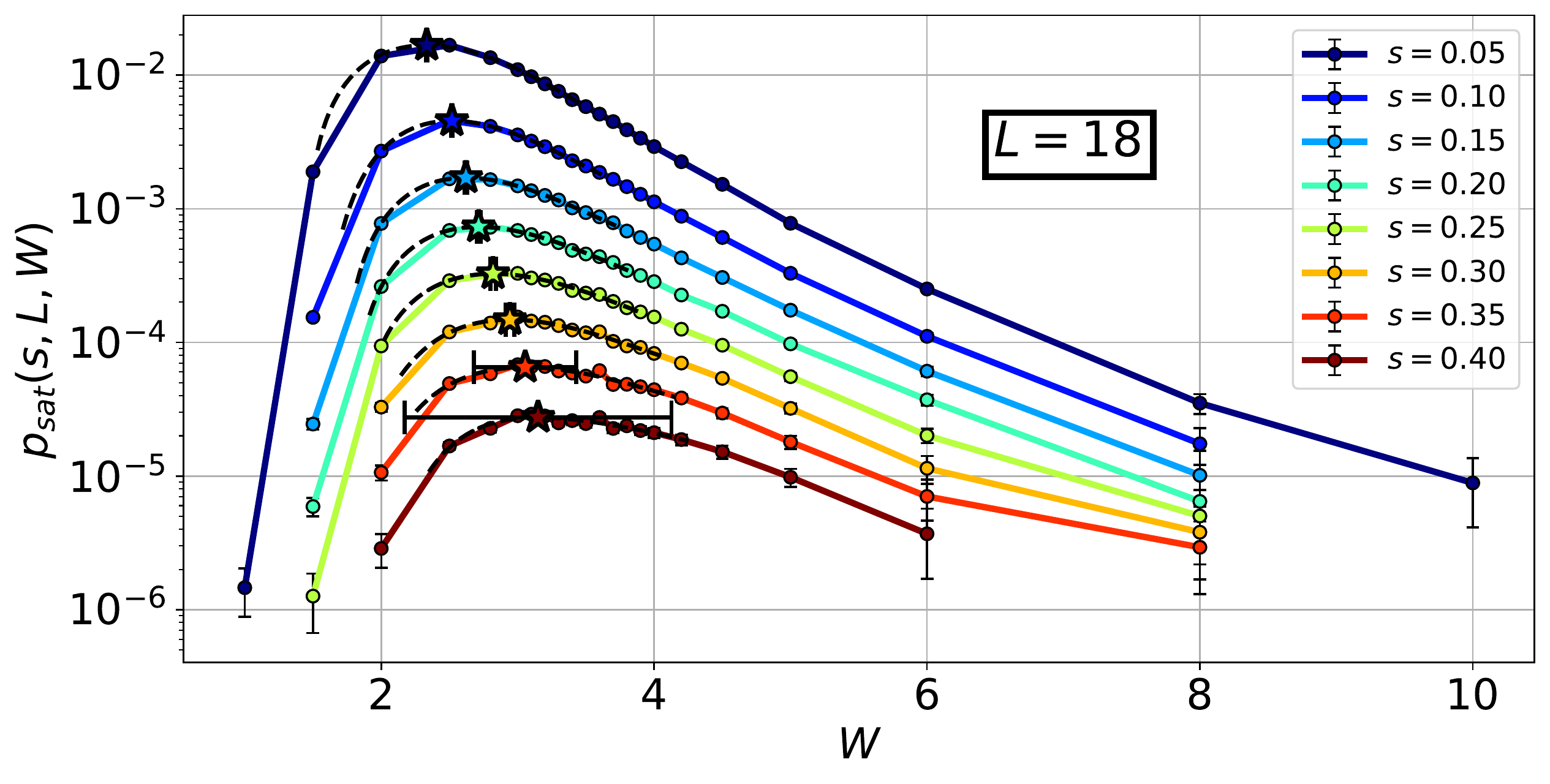}
\includegraphics[width=1.00\columnwidth]{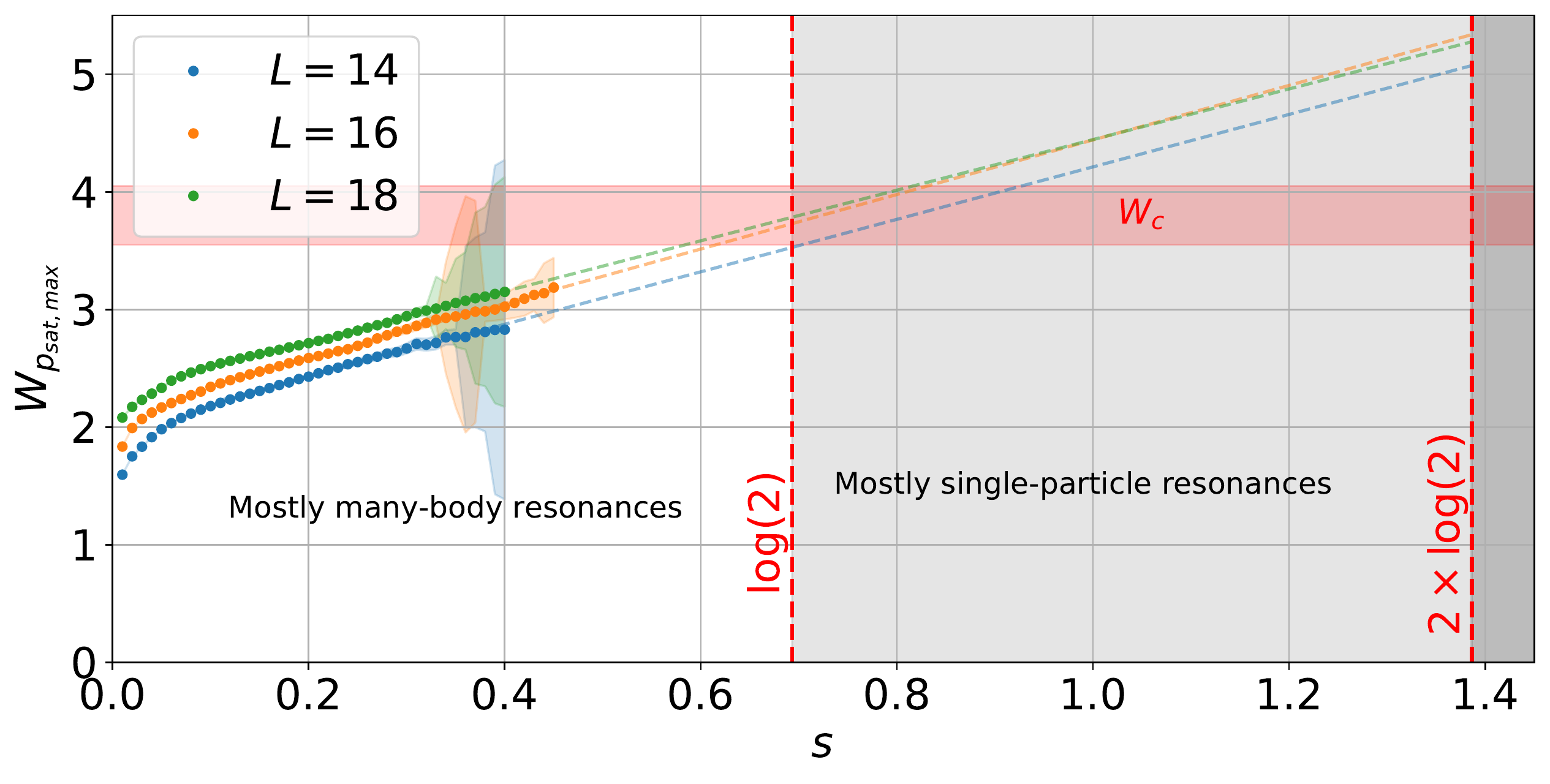}
\caption{\label{fig:p_sat}
\textbf{Top:}  $p_{sat}(s, L, W)$ (see Sec.~\ref{sec:proliferation}) as a function of $W$.
The position of the maxima $W_{max} \equiv W_{p_{sat, max}}(s)$ (stars) is obtained from a local high-order polynomial fit; we find good results fitting to a polynomial of degree seven for all points available in the interval $W \in [0, 6.2]$.
Error bars, both in $p_{sat}$ and $W_{max}$, represent the standard deviations of 200 bootstrap resamples over disorder realizations; the estimation of the $W_{max}$ is highly sensitive across resamples for the largest values of $s$, hence the large error bars.
\textbf{Bottom:} Values of $W_{max}$ for different $L$ and their linear extrapolations for large $s$.
Typically, numerical studies find a critical disorder strength of about $W_c \approx 3.8$; as a visual guide, we have shaded in red the region where $W_c$ is thought to be in the thermodynamic limit.
The extrapolations are compatible with $W_{max} \approx W_c$ when $s = \log(2)$. 
}
\end{figure}

In this section we study the strong tail of the distributions of the $QMI$, \emph{i.e.}, the probability that a pair of sites at range $r$ apart has a very large $QMI$.
In contrast to the typical values of the distribution of Fig.~\ref{fig:histograms_qmi} that were studied earlier in Section~\ref{sec:typical}, we now focus on the upper end of these distributions.
Looking at these extreme values of mutual information is probing rare resonances in the system.

To systematically study the strong tail of the distributions of the $QMI$, we define $p_r(s)$, \emph{i.e.}, the probability that a pair of sites $i$ and $j$ with range $r = |i - j|$ has a $QMI$ larger than a threshold $s$.
While we are mostly interested in $s = \log(2)$ (or $s = 2 \log(2)$) it is impractical to get statistics on these values of $s$ given the rarity of such resonances.
Instead, we consider a set of systematically increasing $s$ out to $s \approx 0.4$. 

Fig.~\ref{fig:prob_qmi} shows the decay of $p_r(s)$ as a function of $r$, for fixed $s$ ($s = 0.4$ in the figure).
We find $p_r(s)$ becomes range invariant around the transition ($W = 3$) for large enough $r$ (but away from the largest values of $r$, in order to avoid finite size effects).
Deep in the MBL phase, our data shows $p_r(s)$ decays with $r$ up to the ranges that we have access to; at more moderate values of $W$, while $p_r(s)$ gets smaller, our data is not sufficient to distinguish between decaying and range-invariant behavior.
In the ergodic phase, $p_r(s)$ decays very rapidly with $r$ leaving us with very few samples to analyze.
That said, the physics of the decay in the ergodic phase are fundamentally different than in the MBL phase.
In the ergodic phase, $p_r(s)$ is small due to every spin being weakly entangled with all other spins, which together with the monogamy of the $QMI$ leaves no room for a strong pairwise mutual information.
In the MBL phase, $p_r(s)$ is small because spins at large $r$ are very rarely entangled with each other.  
This can be seen from Fig.~\ref{fig:histograms_qmi}: while the ergodic distributions of $\log(QMI)$ are compact around a small value of the $QMI$ the MBL distributions are broad and centered around an exponentially small value of the $QMI$, with very rare strong values of the $QMI$ at large range.

\subsection{Proliferation of strong long-range correlations around the transition}
\label{sec:proliferation}

To better quantify the behavior of the strong $QMI$ pairs, we consider the saturation probability $p_{sat}(s,L,W)$ as the probability $p_r(s)$ at large $r$ for each tuple $(s, L, W)$.
The value of $p_{sat}$ is shown in Fig.~\ref{fig:prob_qmi}(right) as a dashed line; the saturation value decays slightly with $L$. 
We extract $p_{sat}(s,L,W)$ from all values of $(W,L)$ where $p_r(s)>0$ at all $r$; the value is extracted by averaging $p_r(s)$ in the interval $r\in [10,L - 1]$.
In practice, this may include some values of $(s,L,W)$ which are not truly saturated; this will not affect the qualitative results we are considering here as we care about the large values of $p_{sat}$ which indeed look convincingly saturated.   

Fig.~\ref{fig:p_sat}(top) presents the values of $p_{sat}$ as a function of $W$ for different thresholds $s$ for a system of size $L = 18$.
We find that the maximum value $W_{max}(s,L)$ of $p_{sat}(s, W, L)$ (shown by a star) is at a disorder strength $W$ close to the transition.
Notice that the position of $W_{max}(s,L)$ becomes unstable at large threshold $s$, due to the small number of samples past that threshold as well as the flatness of the curves around the maximum.

In the bottom panel of Fig.~\ref{fig:p_sat} we plot $W_{max}(s,L)$ as a function of threshold $s$ for different $L$, finding $W_{max}(s,L)$ rises linearly with $s$.
A particularly interesting value of the threshold is $s=\log 2$, which corresponds to the value of the pairwise $QMI$ between all pairs of spins in the canonical multi-site resonating ``cat'' state: ${1 \over \sqrt{2}} (|\Psi_1\rangle + |\Psi_2\rangle)$ where $|\Psi_1\rangle$ and $|\Psi_2\rangle$ are product states which differ in $k>2$ spins.
Although we cannot get any statistics on values of $s$ which are this large, a linear extrapolation finds that $W_{max}(\log 2) \approx 3.8$ at $L=18$, surprisingly close to the best estimates of the transition from scaling collapse~\cite{luitz_many-body_2015}.
This suggests the existence of long-range multi-site resonances at the critical point whose proliferation has been suggested in being responsible for melting MBL; see Ref.~\cite{villalonga_eigenstates_2020} for a complementary numerical approach for probing these long-range resonaces.  

%% file: extreme_entanglement.tex
\section{Extreme entanglement eigenvalues}
\label{sec:entanglement}

In this section, we study $\lambda_2 \equiv -\log(\rho_2)$, where $\rho_2$ is the second singular value of the reduced density matrix of a subsystem over an eigenstate of the Hamiltonian in Eq.~\eqref{eq:model}.
Ref.~\cite{samanta_extremal_2020} argues that the probability of $\lambda_2=\log 2$, \emph{i.e.}, 
\begin{equation}
\label{eq:p_star}
    p^* \equiv \lim_{\lambda_2 \rightarrow \log(2)^+} p(\lambda_2),
\end{equation}
is finite throughout the MBL phase and zero in the ergodic phase, allowing it to be used as an order parameter for the many-body localized phase of matter.
Moreover, the authors of Ref.~\cite{samanta_extremal_2020} find $p^*$ is robust to finite size effects, showing negligible variations across different values of $L$ for small system sizes.
Note that $\log(2)$ is the smallest possible value for $\lambda_2$ and corresponds to a single singlet entangling the subsystem to its environment.
Ref.~\cite{samanta_extremal_2020} studies this in the Gaussian-disordered random Heisenberg model, developing evidence for this conjecture.

In this section, our study differs from Ref.~\cite{samanta_extremal_2020} in three important ways.
Instead of the case of random magnetic fields sampled from a Gaussian distribution, we study the uniform-field case of Eq.~\eqref{eq:model}.
Secondly, while Ref.~\cite{samanta_extremal_2020} considered subsystems of size 5 (see Appendix~\ref{sec:app-EE_LA5}), we consider subsystems of size $L/2$.
Finally, to determine $p^*$, Ref.~\cite{samanta_extremal_2020} looks at the probability density function (PDF) of $\lambda_2$, in order to determine whether $p^*$ is finite or zero as it approaches $\log(2)$.
In our results, we find this limit of the PDF is very sensitive to the choice of bin size.
The finite size of the bins of our histograms were giving the illusion that $p^*$ was finite with an estimated value of $p^*$ which depended on bin-size. 
To alleviate this problem, we instead consider the cumulative distribution function (CDF), which has no binning and which we find presents a more robust method to estimate the behavior of the distribution in the limit of $\lambda_2 \rightarrow \log(2)$.
We then look at the behavior of the CDF  measuring the exponent $\gamma$ of its algebraic approach to $\log(2)$, \emph{i.e.}, 
\begin{align}
\label{eq:cdf}
    \lim_{\lambda_2 \rightarrow \log(2)^+} CDF(\lambda_2) \propto [(\lambda_2 - \log(2))]^\gamma \text{.}
\end{align}
For the PDF of $\lambda_2$ to be non-zero as it approaches $\log 2$ (\emph{i.e.}, $p^* \neq 0$) the CDF has to approach $\log(2)$ with $\gamma = 1$.
On the contrary, $\gamma > 1$ implies that $p^* = 0$.

\begin{figure}
\centering
\includegraphics[width=1.00\columnwidth]{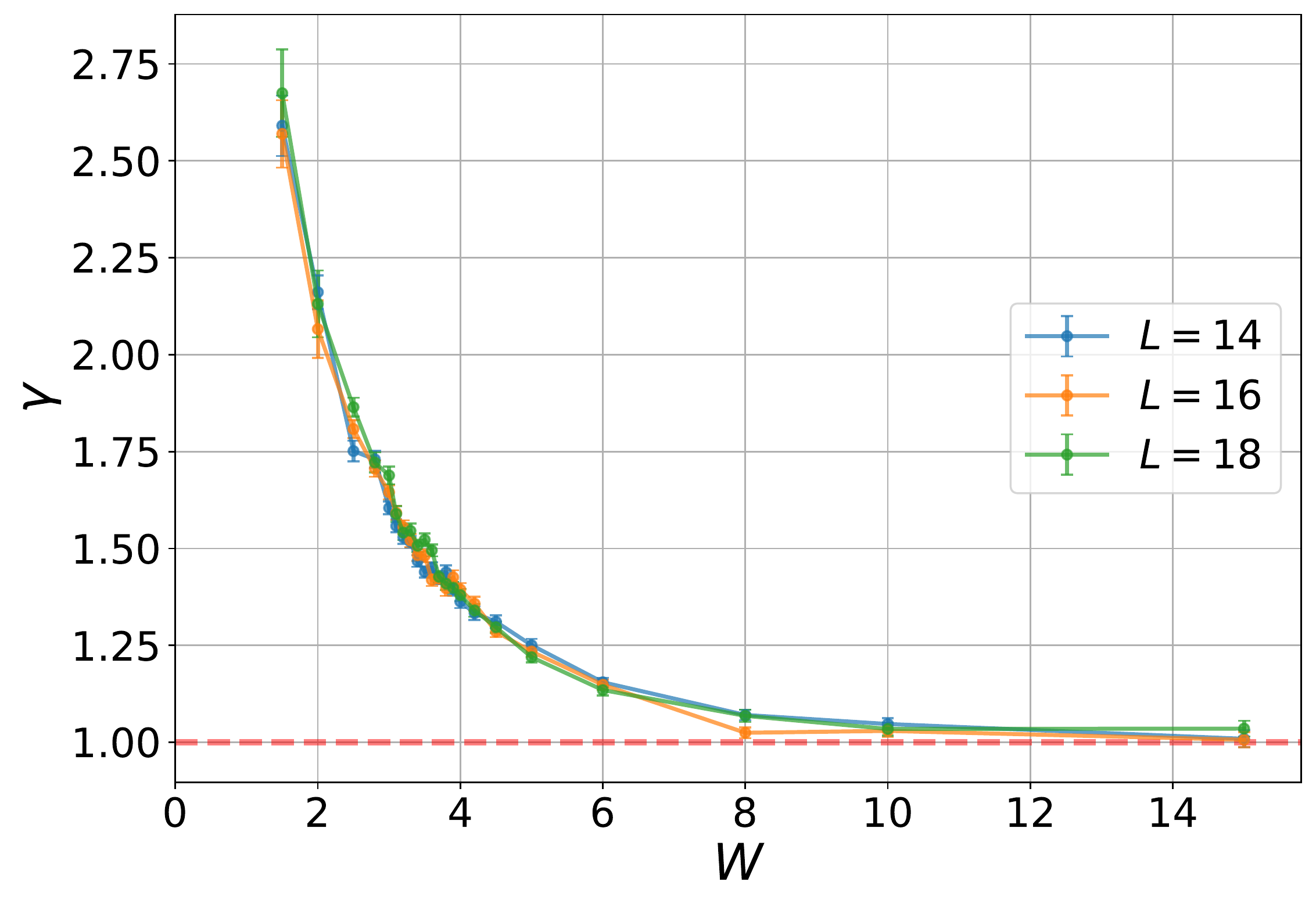}
\caption{\label{fig:gamma} Exponent $\gamma$ of  Eq.~\eqref{eq:cdf}.
Confidence intervals correspond to the standard deviation of $\gamma$ from 200 bootstrap resamples over the original disorder realizations.
$\gamma = 1$ deep in the MBL phase, which is compatible with a non-zero value of $p^*$ (see Eq.~\eqref{eq:p_star}) and hence a non-zero probability of finding a single singlet across the half cut in the chain.
At moderate disorder strengths and close to the transition $\gamma > 1$, which implies a vanishing probability of finding a singlet across the chain.
Deep in the ergodic phase we do not have enough extremal data to fit a power law close to $\log(2)$.
See Appendix~\ref{sec:app-lambda_2_fits} for more details on the extraction of $\gamma$ and its error bars. }
\end{figure}

Fig.~\ref{fig:gamma} shows the empirical values of $\gamma$ we find. 
Deep in the MBL phase (i.e. $W >8$), we find $\gamma \approx 1 $, compatible with $p^*>0$ at large $W$ (although we can not rule out that even there $\gamma$ is still marginally above 1); therefore, we potentially expect a single singlet spanning the two halves of the system deep in the MBL phase.
At moderate values of $W$ but still in the MBL phase as well as at the transition, $\gamma$ is significantly above 1, indicating that $p^* = 0$. 
Like Ref.~\cite{samanta_extremal_2020}, we do find that $\gamma$ shows low sensitivity to small changes in system size.
This is presumably due to the fact that singlets across a cut entangle primarily, at moderate and strong disorder, spins close to the cut, which are far away from the boundaries of the system and therefore have low sensitivity to its size (see Appendix~\ref{fig:app-gamma_LA5}).

%% file: conclusion.tex
\section{Conclusions}
\label{sec:conclusions}

Most of this work focuses on the sutdy of the correlations throughout the ergodic-MBL phase diagram of the model of Eq.~\ref{eq:model} through the distributions of the logarithm of the $QMI$.
We have focused on two aspects of this distribution: the overall shape of the distribution (\emph{i.e} its moments) and the extreme tails of the distribution.
Given the large amount of data ($10^6$ eigenstates at energy density $\epsilon \approx 0.5$ per $(W, L)$) and relatively large values of $L$, we can get precise results on multiple aspects of these distributions.

The main contribution of our work is identifying a region at moderate disorder strength that shows a stretched exponential decay ($QMI_{typ} = e^{-A r^\beta}$) of the typical correlations (log-averaged $QMI$) as a function of range $r$.
At the transition, this decay takes the form $QMI_{typ} = e^{-A \sqrt{r}}$, which is also found in the random singlet phase.
To further under the universality of the distribution of $\log(QMI)$ at the transition we consider higher moments.
We find the $\log(QMI)$ is universal for all moments except for the second moment, which scales quadratically with range ($\sigma[\log(QMI)]$ scales linearly).
This has some similarities with the distribution of the logarithm of the correlations in the random singlet phase which is conjectured to be universal after rescaling by $\sqrt{r}$ to remove its mean~\cite{fisher_random_1994} (note that rescaling by its mean does not affect the scaling of the standardized moments).
The distributions of the $\log(QMI)$ seem therefore to satisfy a weaker version of the universality conjectured for the random singlet phase.
This aspect of the distributions should be a key feature in determining the universality class of the MBL-ergodic transition and can serve as a constraint for the panoply of RG results which attempt to explain the phenomenology of the MBL-ergodic transition.

A second key result has involved studying the atypically strong correlations across the system.
Our results show the existence of large range-invariant pair-wise $QMI$ at the transition suggesting the existence of proliferating multi-site resonances. 

Finally, we have also studied the extremal values of the second entanglement eigenvalue, $\lambda_2$, which signals the presence of a single bit of entanglement across a bipartition of the system.
Our results show that this probability is only finite deep in the MBL phase, and vanishes as a power law at moderate values of the disorder strength and in the ergodic phase.

%% file: acknowledgments.tex
\begin{acknowledgments}

We would like to acknowledge useful discussions with Vedika Khemani, Anushya Chandran, Romain Vasseur, Chris Laumann, Lo\"ic Herviou, Greg Hamilton, and Eli Chertkov.
We both acknowledge support from the Department of Energy grant DOE de-sc0020165.
BV also acknowledges support from the Google AI Quantum team.
This project is part of the Blue Waters sustained petascale computing project, which is supported by the National Science Foundation (awards OCI-0725070 and ACI-1238993) and the State of Illinois.
Blue Waters is a joint effort of the University of Illinois at Urbana-Champaign and its National Center for Supercomputing Applications.

\end{acknowledgments}

%% file: appendix.tex
\section{Computation of the $QMI$}
\label{sec:app-qmi}

The two-site $QMI$ for a spin-$1 \over 2$ total magnetization preserving model like the one in Eq.~\ref{eq:model} is computed from a few expectation values: $\langle n_i \rangle$, $\langle n_j \rangle$, $\langle \sigma_i^+ \sigma_j^- \rangle$, and $\langle n_i n_j \rangle$, where $n_i \equiv \frac{\sigma_i^z + 1} {2}$.
In particular,
\begin{align}
    QMI_{ij} = S_i + S_j - S_{ij} \text{,}
\end{align}
where
\begin{align}
    \label{eq:s_i}
    S_i = &-\langle n_i \rangle \log(\langle n_i \rangle) - (1- \langle n_i \rangle) \log(1 - \langle n_i \rangle) \\
    \label{eq:s_ij}
    S_{ij} = &-\langle n_i n_j \rangle \log\langle n_i n_j \rangle \nonumber \\
    &- \langle(1 - n_i) (1 - n_j)\rangle \log \left[\langle(1 - n_i) (1 - n_j) \rangle \right] \nonumber \\
    &- \lambda_+ \log(\lambda_+) \nonumber \\
    &- \lambda_- \log(\lambda_-) \nonumber \\
\end{align}
with
\begin{align}
    \label{eq:lambda}
    \lambda_{ij, \pm} = \frac{\langle ( n_i - n_j )^2 \rangle \pm \sqrt{\left(\langle n_i \rangle - \langle n_j \rangle \right)^2  + 4 | \langle\sigma_i^+ \sigma_j^-\rangle |^2}}{2}
\end{align}
with
\begin{align}
    \langle(1 - n_i) (1 - n_j)\rangle &= 1 - \langle n_i \rangle - \langle n_j \rangle + \langle n_i n_j \rangle  \\
    \langle (n_i - n_j)^2 \rangle &= \langle n_i \rangle + \langle n_j \rangle - 2 \langle n_i n_j \rangle \text{.}
\end{align}

\section{Numerical precision in the computation of the $QMI$}
\label{sec:app-precision}

The eigenstates of the model of Eq.~\ref{eq:model} are obtained with double-precision floating-point numbers, which means that the largest vector entry (assuming it is $\mathcal{O}(1)$) has a precision of $\approx 10^{-15}$.
This implies the expectation values of Appendix~\ref{sec:app-qmi} will also have a precision of about $10^{-15}$.
When computing the $QMI$ from these expectation values, there are two points where the precision might drop.
First, all terms in Eqs.~\ref{eq:s_i} and \ref{eq:s_ij} are of the form $x \log(x)$, with $0 \leq x \leq 1$.
A careful inspection shows that the precision of this quantity (if $x$ has a precision of $10^{-15}$) is of order $10^{-14}$ (\emph{e.g.}, $x \log(x) = -3.45 \cross 10^{-14}$ for $x = 10^{-15}$), and the $QMI$ overall has a precision of about $10^{-14}$.
Starting from vectors of precision $10^{-15}$ (as we do), there is nothing we can do about this drop in precision.

There is another point at which precision can drop, \emph{i.e.}, the computation of $\lambda_{\pm}$, which involves squaring expectation values that we have obtained with a precision of $10^{-15}$ followed by a square root.
The square needs of a precision of $10^{-30}$ in order to keep an overall $10^{-15}$ after the square root is taken.
The use of slightly higher-precision floating-point types such as \texttt{numpy.longdouble} or \texttt{numpy.float128} in \texttt{python} does not solve this issue (note these types have typically a precision of $10^{-18}$).
We instead make use of the \texttt{decimal} module in \texttt{python} in order to work with arbitrary precision; in practice we use 60 decimal places.
This gives us confidence that $x$ has precision $10^{-15}$ in the $x \log(x)$ terms (note we cannot improve this precision, given that we start our computation from double-precision vector entries).
The $QMI$ ultimately has a precision of $\approx 10^{-14}$.
Below this threshold, the distributions of the $QMI$ of Fig.~\ref{fig:histograms_qmi} still look smooth, but should not be trusted.
In practice we consider only those distributions which have at least 99\% of their mass above $10^{-14}$, \emph{i.e.}, an order of magnitude above the typical double-precision threshold of $10^{-15}$.

Finally, we find in practice that $1 - \langle n_i \rangle$ and $\langle(1 - n_i) (1 - n_j)\rangle$ (see Appendix~\ref{sec:app-precision}) are on rare occasions negative and of order $\lessapprox 10^{-15}$.
This is expected from the precision we work with.
Since this is a problem for the evaluation of their corresponding logarithms, we substitute these values by $10^{-20}$.
Note that, given the magnitude of these terms, this substitution does not reduce the precision of the $QMI$ further.

\section{Alternative view of $p_{sat}$ and its maxima}
\label{sec:app-maxima_p_sat}

\begin{figure}[t]
\centering
\includegraphics[width=1.00\columnwidth]{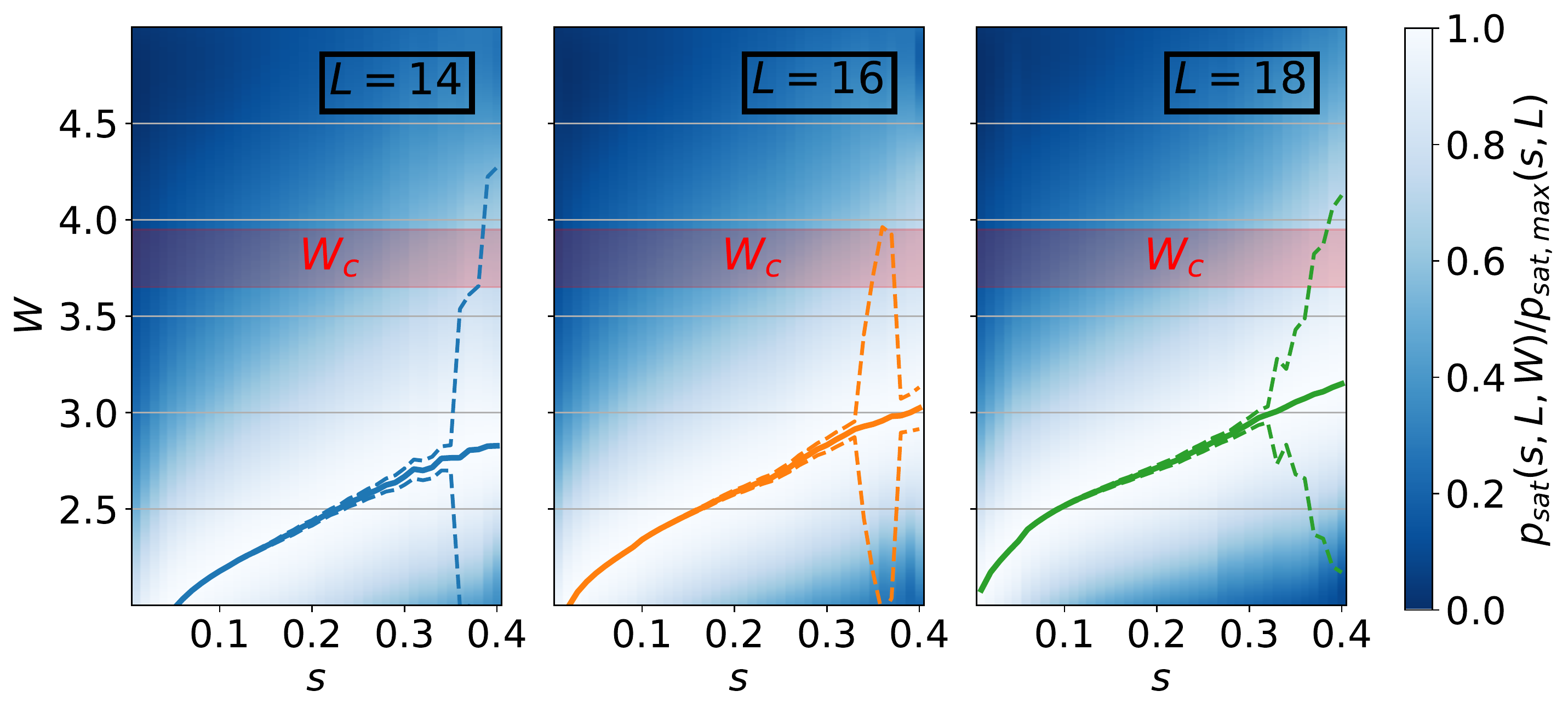}
\caption{\label{fig:app-p_sat}
Colormap with the high-order polynomial fits of the top panel, normalized by the maximum value of $p_{sat}$.
$W_{p_{sat, max}}(s)$ is plotted, which is compatible with the maximum probability of finding strong, long-range $QMI$ bonds at the critical disorder strength $W_c$ in the thermodynamic limit.
}
\end{figure}

Fig.~\ref{fig:app-p_sat} shows a colormap with all fitted curves of $p_{sat}(s, W, L)$ obtained for thresholds between $s = 0$ and $s = 0.4$ and for three different system sizes.
The curves shown in the colormap have been normalized by their maximum value, $p_{sat, max}$, so that they are all visible in the plot.

\section{Second entanglement eigenvalue for a susbsystem of size 5}
\label{sec:app-EE_LA5}

\begin{figure}
\centering
\includegraphics[width=1.00\columnwidth]{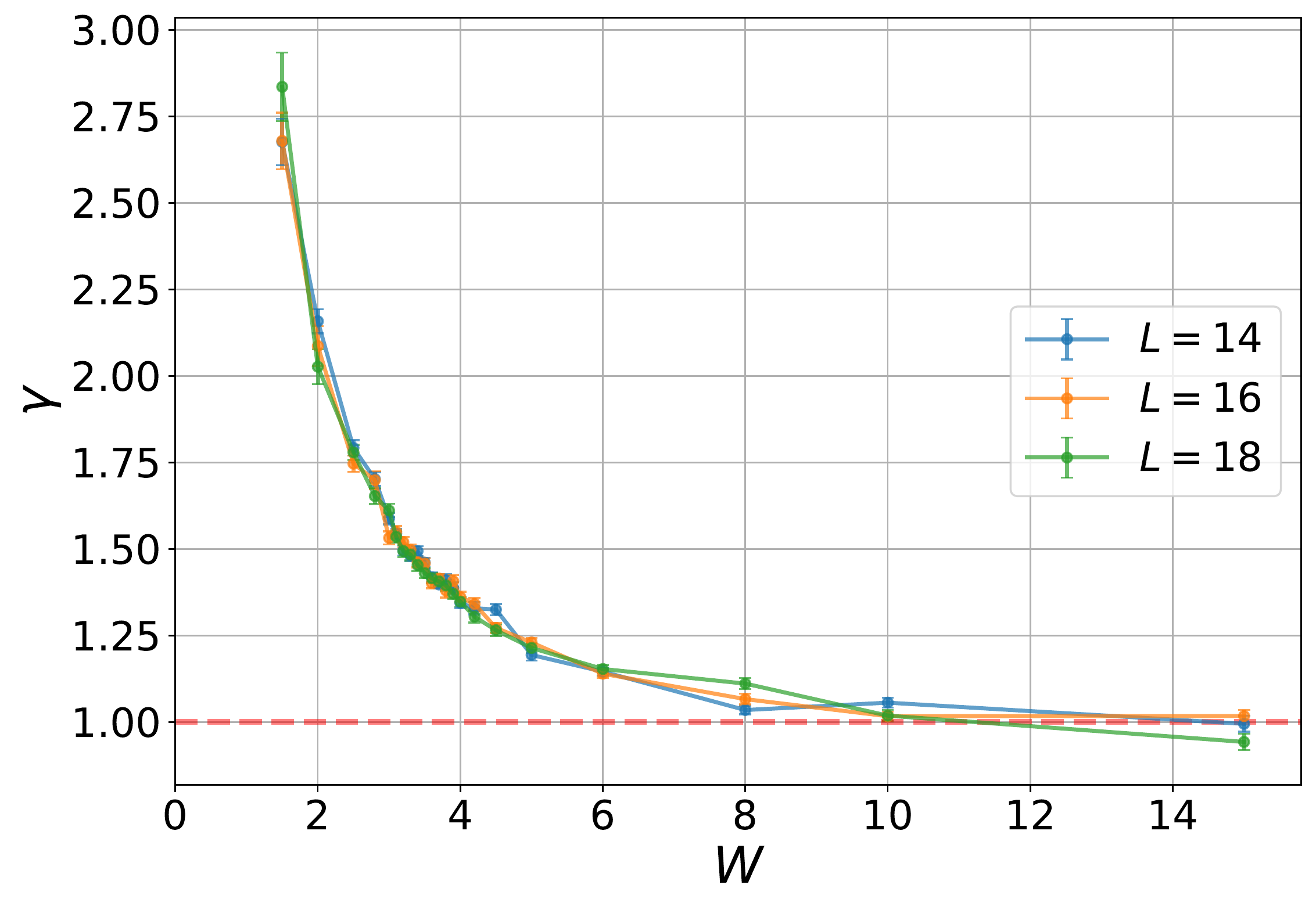}
\caption{\label{fig:app-gamma_LA5}
Same as Fig.~\ref{fig:gamma} of the main text for a cut between a subsystem $A$ of size $L_A = 5$ and the rest of the chain.
Error bars represent the standard deviation of $\gamma$ across 200 bootstrapping resamples over disorder realizations.
}
\end{figure}

In Section~\ref{sec:entanglement} we studied the extremal values of the second bipartite entanglement eigenvalue $\lambda_2$ over a cut at the middle of the chain.
However, Ref.~\cite{samanta_extremal_2020} studied this quantity over a cut between a subsystem $A$ of size $L_A = 5$ and the rest of the chain.
In Fig.~\ref{fig:app-gamma_LA5} we present the values of $\gamma$ (see Section~\ref{sec:entanglement}) for such a cut.
There is no substantial difference between this cuts and the half-cut of the entanglement entropy.

\section{Log-log plots $QMI_{typ}$ for the extraction of $\beta$}
\label{sec:app-beta_fits}

\begin{figure*}[t]
\centering
\includegraphics[width=2.00\columnwidth]{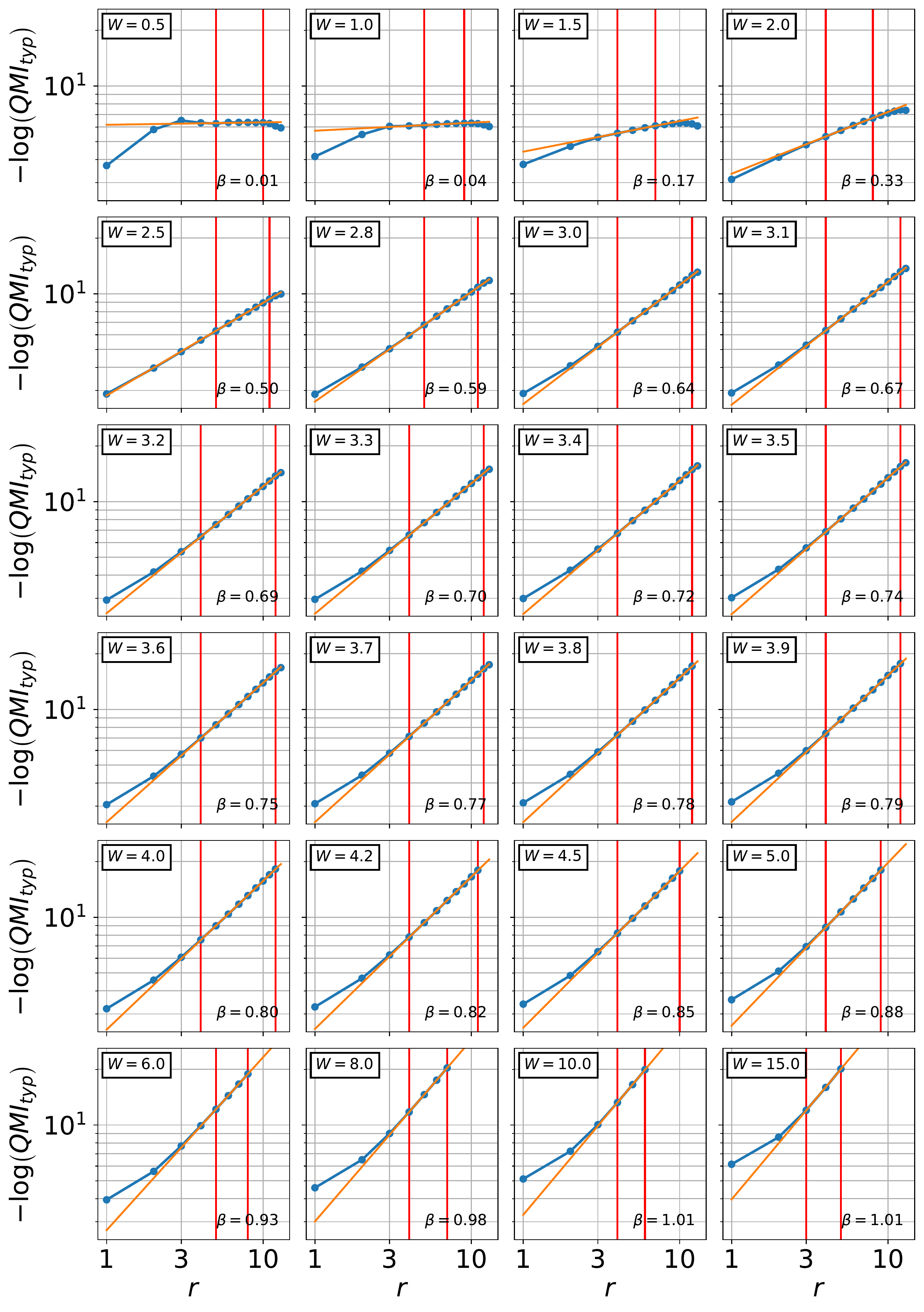}
\caption{\label{fig:beta_fits_L14} 
Log-log plots of $-\log_{10}(QMI_{typ})$ as a function of range $r$ for $L = 14$.
We extract the exponent $\beta$ of the stretched exponential of Section~\ref{sec:typical} from the slope of a linear fit in the interval given by the red vertical lines.
We can see that the fit is good at strong disorder and down to the system size dependent value of $W$ for which $\beta = 0.5$, corresponding to the ergodic-MBL transition.
$W_{1 / 2} = 2.50$ for $L = 14$; we can see that the fit is particularly good around $W_{1 / 2}$, even at the lowest ranges $r$.
At $W < W_{1 / 2}$ the linear fit overestimates $\log(QMI_{typ})$ at low $r$, while at $W > W_{1 / 2}$ the linear fit underestimates $\log(QMI_{typ})$ at low $r$.
}
\end{figure*}

\begin{figure*}[t]
\centering
\includegraphics[width=2.00\columnwidth]{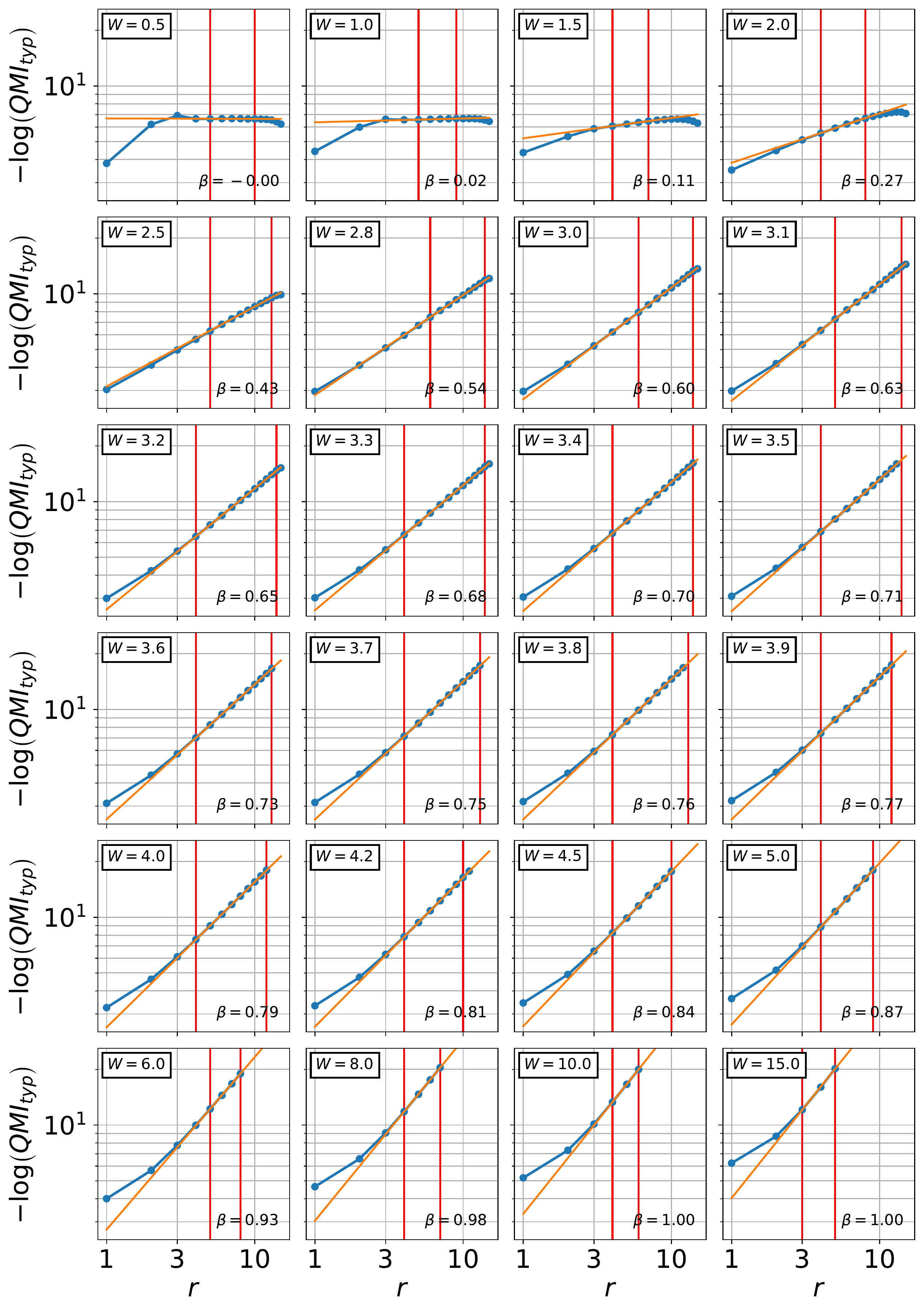}
\caption{\label{fig:beta_fits_L16} Same as Fig.~\ref{fig:beta_fits_L14} for systems of size $L = 16$.
$W_{1 / 2} = 2.69$ for $L = 16$.
}
\end{figure*}

\begin{figure*}[t]
\centering
\includegraphics[width=2.00\columnwidth]{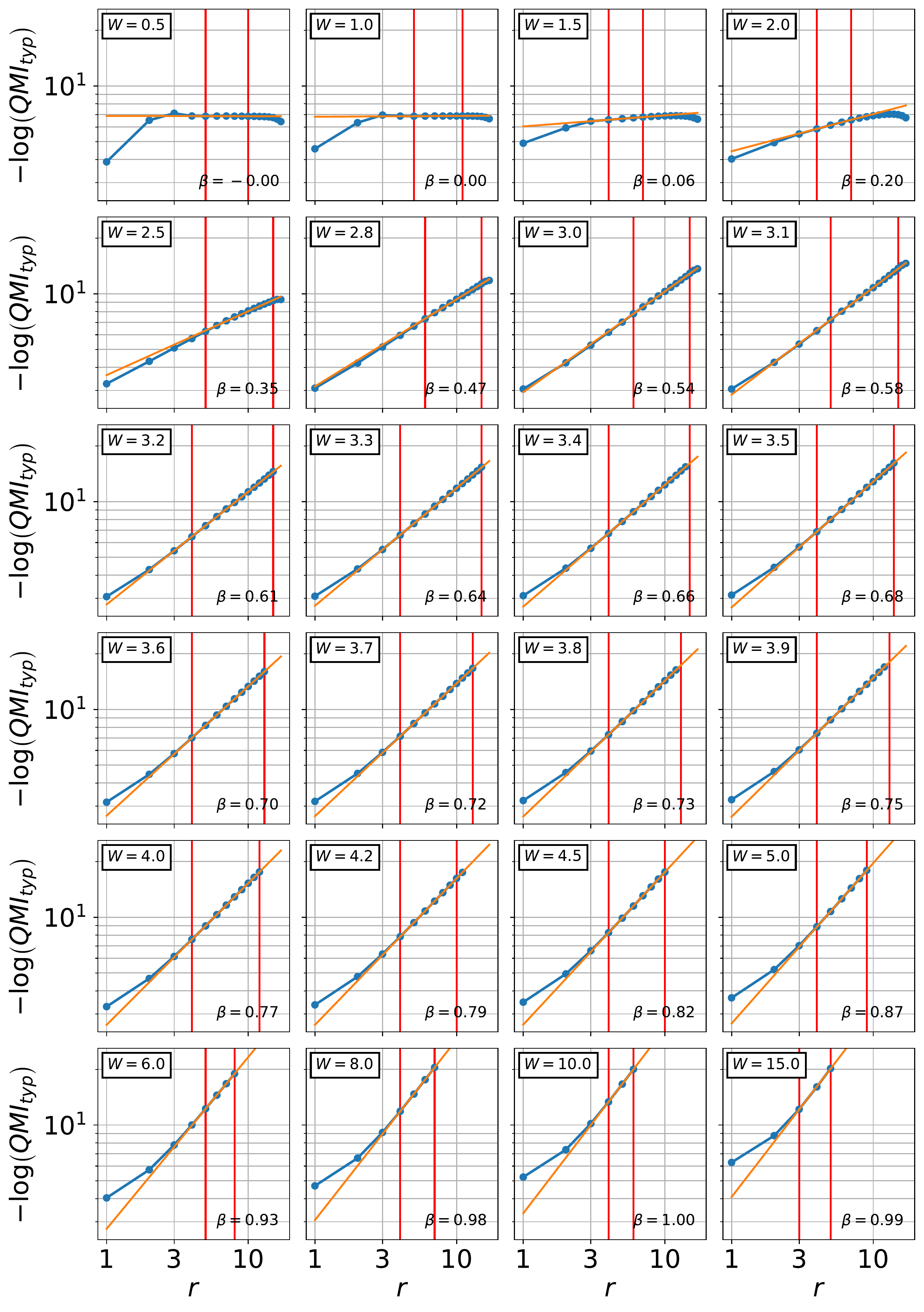}
\caption{\label{fig:beta_fits_L18} Same as Fig.~\ref{fig:beta_fits_L14} for systems of size $L = 18$.
$W_{1 / 2} = 2.88$ for $L = 18$.
}
\end{figure*}

Figs.~\ref{fig:beta_fits_L14}, \ref{fig:beta_fits_L16}, and \ref{fig:beta_fits_L18} present all linear fits of $\left\langle \log(QMI) \right\rangle$ as a function of range $r$ in a log-log scale for all $(L, W)$.
As discussed in Section~\ref{sec:means}, this lets us extract the exponent $\beta$ of the stretch exponential decay of $QMI_{typ} = e^{-A r^\beta}$.

\section{Standard deviations of $\log(QMI)$ and linear fits to extract $C$}
\label{sec:app-C_fits}

\begin{figure*}[t]
\centering
\includegraphics[width=2.00\columnwidth]{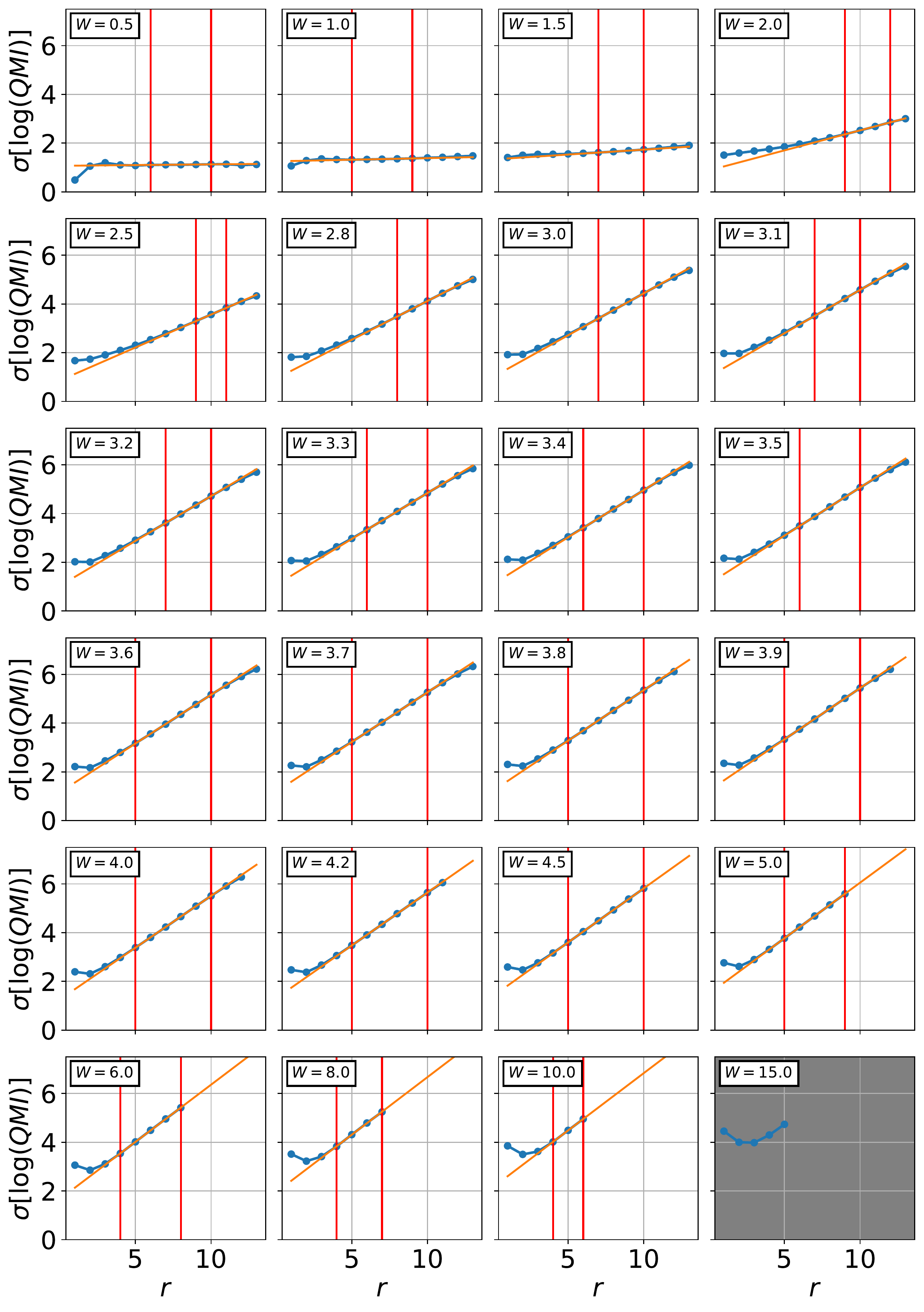}
\caption{\label{fig:std_fits_L14} 
Linear fits to the standard deviation of $\log(QMI)$ as a function of $r$ for different values of $W$ and a system of size $L = 14$.
The linear fits are in general of very good quality, with the exception of $1 \lessapprox W \lessapprox 2$ and $W \gtrapprox 8$, where we only have a few points in the linear scaling regime; at $W = 15$ we do not have enough points to even enter the linear scaling regime.
}
\end{figure*}

\begin{figure*}[t]
\centering
\includegraphics[width=2.00\columnwidth]{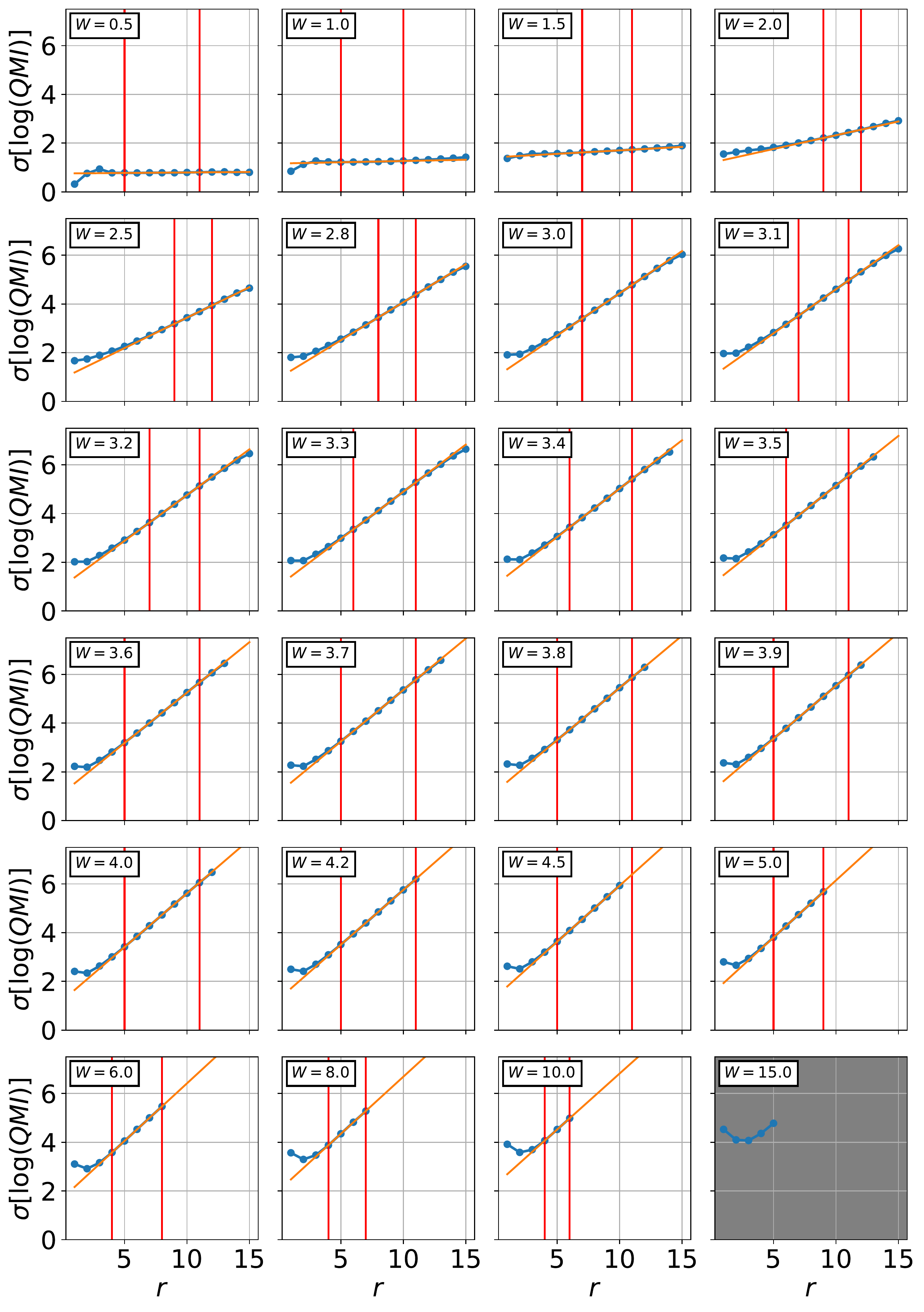}
\caption{\label{fig:std_fits_L16} Same as Fig.~\ref{fig:std_fits_L14} for systems of size $L = 16$.
}
\end{figure*}

\begin{figure*}[t]
\centering
\includegraphics[width=2.00\columnwidth]{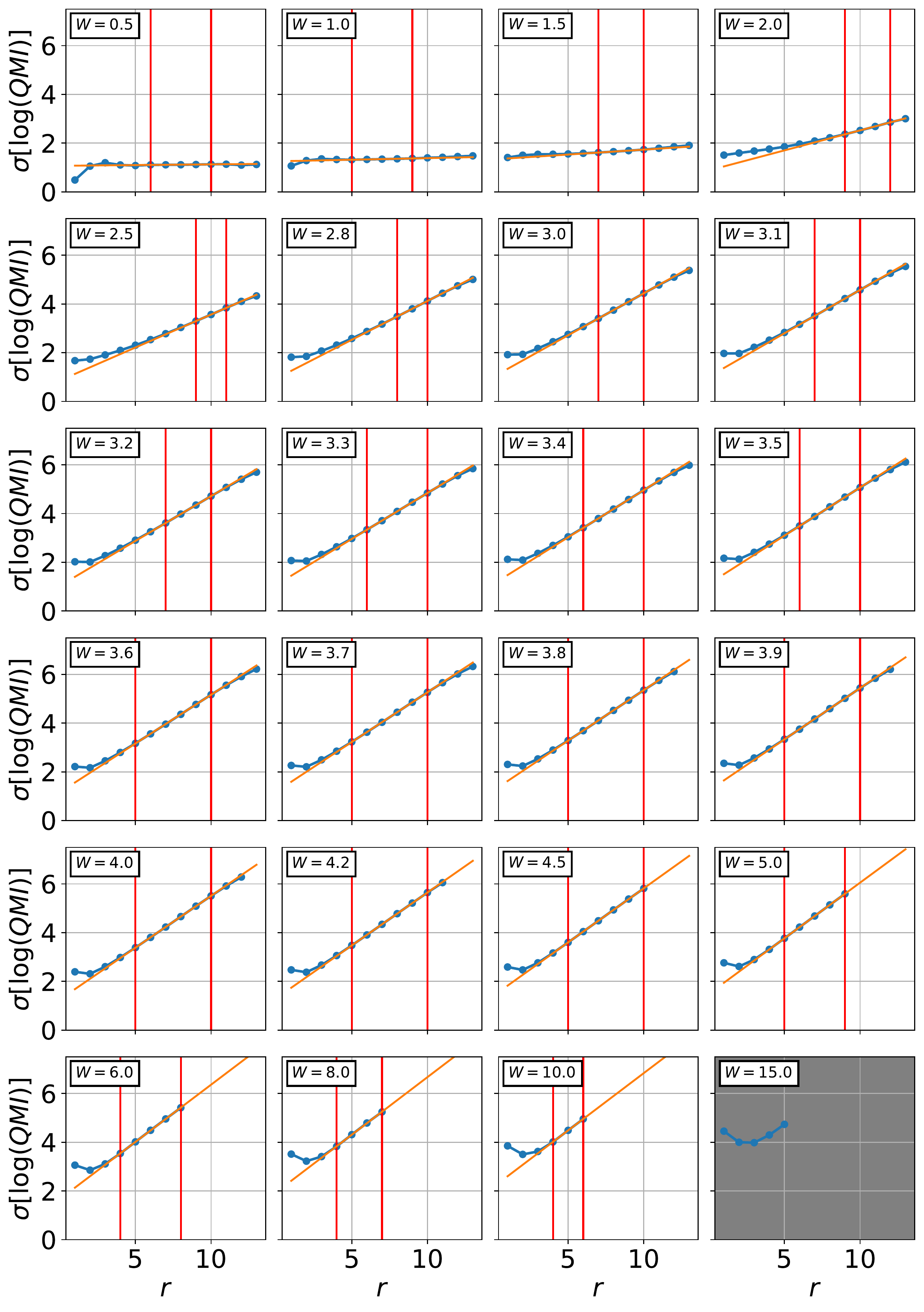}
\caption{\label{fig:std_fits_L18} Same as Fig.~\ref{fig:std_fits_L14} for systems of size $L = 18$.
}
\end{figure*}

Figs.~\ref{fig:std_fits_L14}, \ref{fig:std_fits_L16} and \ref{fig:std_fits_L18} show all linear fits used in Section~\ref{sec:std} in order to extract the slope $C$ of the linear scaling of $\sigma\left[ \log(QMI) \right] = C r + D$ as a function of range $r$.

\section{Linear fits to the CDF of the second bipartite entanglement eigenvalue}
\label{sec:app-lambda_2_fits}

\begin{figure*}[t]
\centering
\includegraphics[width=2.00\columnwidth]{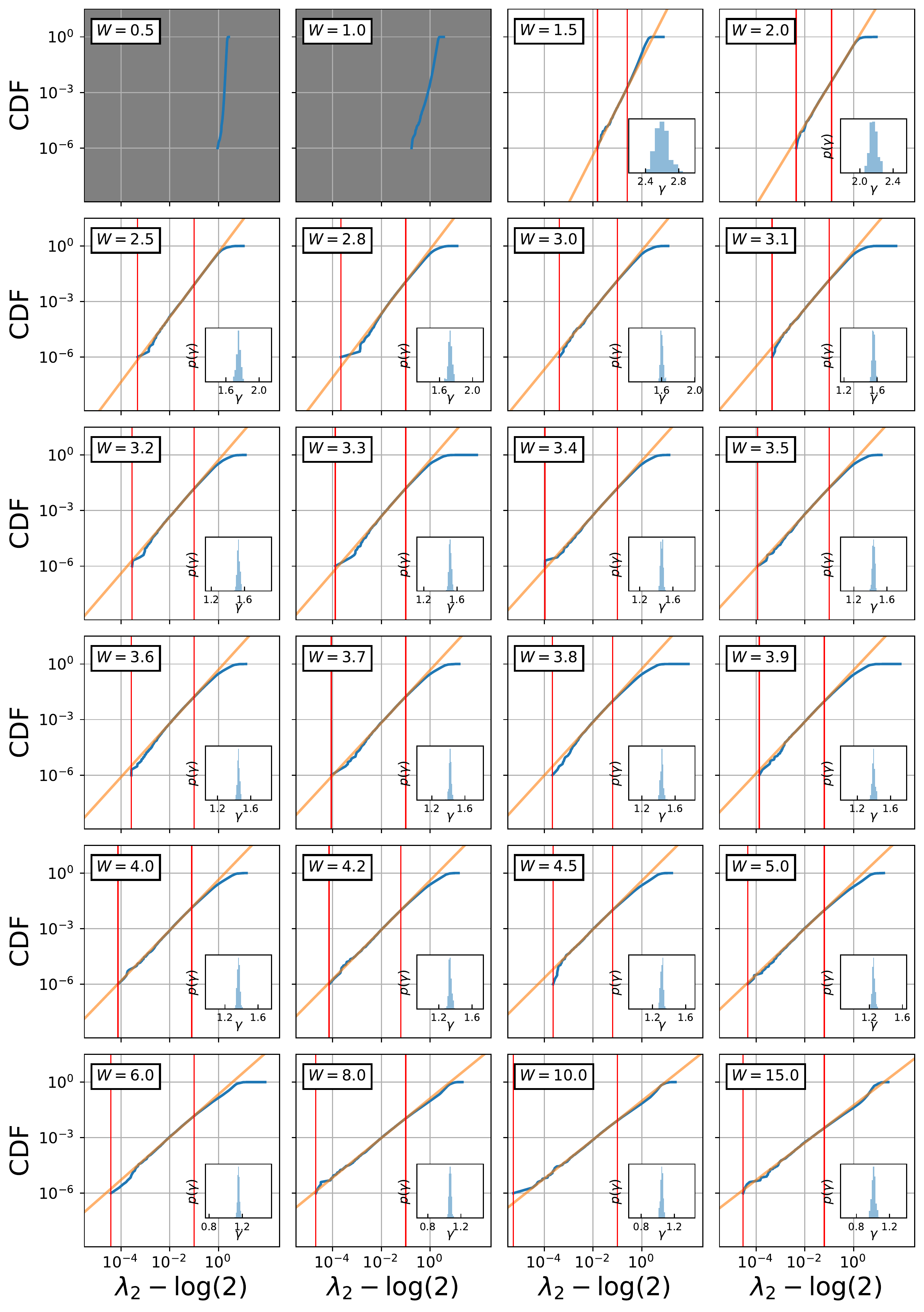}
\caption{\label{fig:lambda_2_fits_L14}
Linear fits of the log-log representation of the left-side tail of the CDF of the second eigenvalue of the bipartite reduced density matrix of a system of size $L = 14$.
Red vertical lines denote the ends of the interval that contains the points used in each case to make the fit.
The slope of the fit gives us the exponent $\gamma$ in the expression $CDF(\lambda_2) \approx (\lambda_2 - \log(2))^\gamma$, which is equivalent to equivalently $PDF(\lambda_2) \approx (\lambda_2 - \log(2))^{\gamma - 1}$ up to an additive constant.
Below $W = 1.5$, the low $\lambda_2$ data is inexistent and we do not attempt to extract an exponent $\gamma$ from a linear fit.
\textbf{Inset:} probability distribution of $\gamma$ extracted from 200 bootstrap resamples over the 200K disorder realizations for this system size.
Bootstrapping was used in order to compute confidence intervals for $\gamma$, shown in Fig.~\ref{fig:gamma}.
}
\end{figure*}

\begin{figure*}[t]
\centering
\includegraphics[width=2.00\columnwidth]{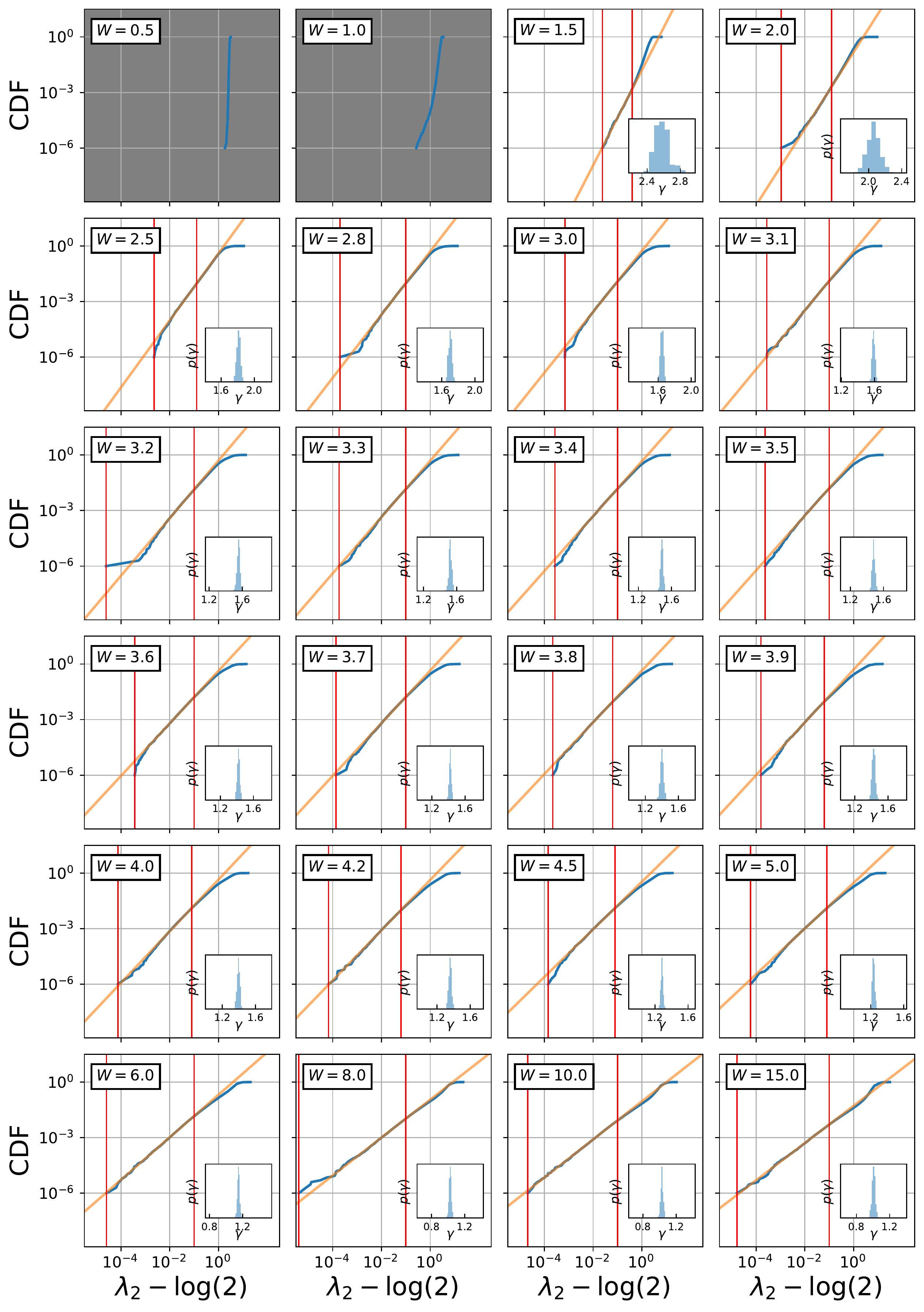}
\caption{\label{fig:lambda_2_fits_L16} Same as Fig.~\ref{fig:lambda_2_fits_L14} for systems of size $L = 16$.
\textbf{Inset:} probability distribution of $\gamma$ extracted from 200 bootstrap resamples over the 200K disorder realizations for this system size.
}
\end{figure*}

\begin{figure*}[t]
\centering
\includegraphics[width=2.00\columnwidth]{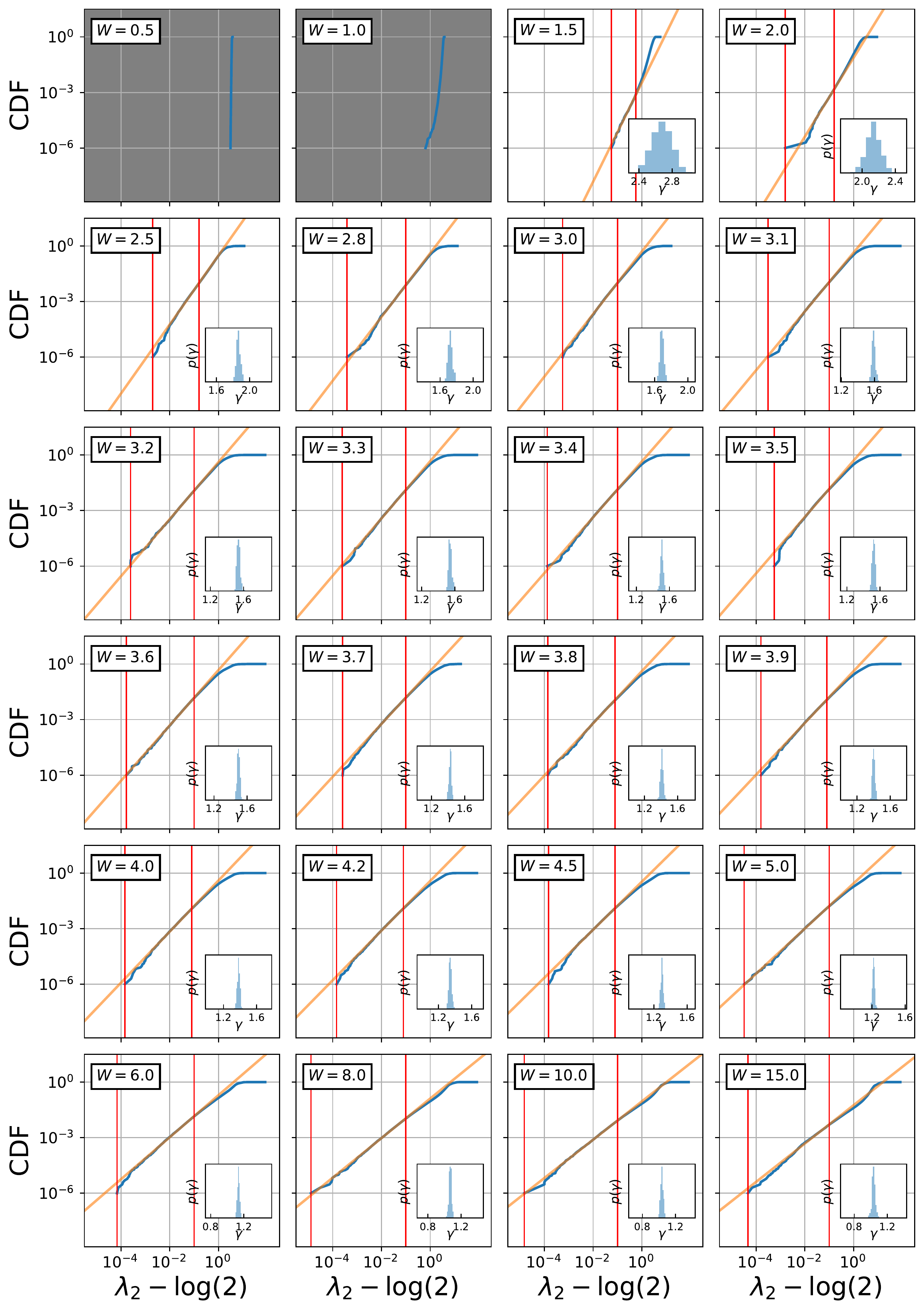}
\caption{\label{fig:lambda_2_fits_L18} Same as Fig.~\ref{fig:lambda_2_fits_L14} for systems of size $L = 18$.
\textbf{Inset:} probability distribution of $\gamma$ extracted from 200 bootstrap resamples over the 10K disorder realizations for this system size.
}
\end{figure*}

Here we present data related to the extraction of the exponent $\gamma$ for the CDF of the second entanglement eigenvalue, $\lambda_2$, when it approaches $\lambda_2 \rightarrow \log(2)^+$, which was discussed in Section~\ref{sec:entanglement}.
In particular, Figs.~\ref{fig:lambda_2_fits_L14}, \ref{fig:lambda_2_fits_L16}, and \ref{fig:lambda_2_fits_L18} linear regression results of the fit on a log-log plot to the left-side tails of $p(\lambda_2)$ as a function of $\lambda_2 - \log(2)$.
The slope of this fit is equal to $\gamma$.
In order to estimate the error bars for $\gamma$, we perform a bootstrapping analysis with 200 resamples over disorder realizations.
The inset of the figures provides the distribution of $\gamma$ from bootstrapping; its standard deviation is taken as an estimate of the error on the estimation of $\gamma$.

